\documentclass[conference]{IEEEtran}
\usepackage{amsmath,amssymb,amsfonts}
\usepackage{microtype}
\usepackage{import}
\usepackage{calc}
\usepackage{cite}
\usepackage{graphicx}
\graphicspath{{./figure/block/}{./figure/plot}}
\usepackage{subcaption}
\usepackage[abbreviations = true]{siunitx}
\usepackage{balance}
\usepackage[all]{nowidow}

\begin{document}
\title{Electromagnetic Signal Injection Attacks on \\ Differential Signaling}

\author{\IEEEauthorblockN{Anonymous Author(s)}}
\author{\IEEEauthorblockN{Youqian Zhang}
\IEEEauthorblockA{University of Oxford\\
youqian.zhang@cs.ox.ac.uk}
\and
\IEEEauthorblockN{Kasper Rasmussen}
\IEEEauthorblockA{University of Oxford\\
kasper.rasmussen@cs.ox.ac.uk}}


\maketitle

\begin{abstract}
  Differential signaling is a method of data transmission that uses two complementary electrical signals to encode information. This allows a receiver to reject any noise by looking at the difference between the two signals, assuming the noise affects both signals in the same way. Many protocols such as USB, Ethernet, and HDMI use differential signaling to achieve a robust communication channel in a noisy environment. This generally works well and has led many to believe that it is infeasible to remotely inject attacking signals into such a differential pair. In this paper we challenge this assumption and show that an adversary can in fact inject malicious signals from a distance, purely using common-mode injection, i.e., injecting into both wires at the same time. We show how this allows an attacker to inject bits or even arbitrary messages into a communication line. Such an attack is a significant threat to many applications, from home security and privacy to automotive systems, critical infrastructure, or implantable medical devices; in which incorrect data or unauthorized control could cause significant damage, or even fatal accidents.

  We show in detail the principles of how an electromagnetic signal can bypass the noise rejection of differential signaling, and eventually result in incorrect bits in the receiver. We show how an attacker can exploit this to achieve a successful injection of an arbitrary bit, and we analyze the success rate of injecting longer arbitrary messages. We demonstrate the attack on a real system and show that the success rate can reach as high as $90\%$. Finally, we present a case study where we wirelessly inject a message into a Controller Area Network (CAN) bus, which is a differential signaling bus protocol used in many critical applications, including the automotive and aviation sector.
\end{abstract}

\section{Introduction}
\label{sec:introduction}

Electrical cables are widely used to wire up devices, enabling signal transmission from one to the other. 
For example, a Universal Serial Bus (USB) cable connects a keyboard to a computer and a High Definition Multimedia Interface (HDMI) cable projects video streams from a laptop to a TV.
Despite the fact that metal conductors in the cables are specifically used for signal transmissions, they can also act like antennas~\cite{balanis2015antenna}, i.e., they will capture environmental electromagnetic interference, which is generated by other electrical devices such as fluorescent lights and motors.
Such captured external interference will constitute a threat to the signal integrity. 
To make the cables less susceptible to external interference, many communication protocols such as USB, HDMI, Ethernet, and Controller Area Network (CAN) are built on the principle of \emph{differential signaling}.

In differential signaling, information is transmitted by a pair of signals that carry equal magnitude but opposite polarities, and each in its own wire.
The information at the receiver is interpreted as the difference between the differential pair of signals. 
Note that the two wires are identical and put close to each other (e.g., twisted cables), implying that they are identical antennas.
If external interference tends to impact the wires, it modifies the differential pair of signals equally, but does not impact the difference leaving the intended information intact.
This is why differential signaling can reject the external interference.
In practice, within the operating frequency range of the receiver, such a rejection ability can attenuate the external interference by more than $\SI{70}{\dB}$ typically~\cite{devices2009op}.
However, the rejection ability deteriorates drastically beyond the operational frequency range~\cite{razavi2002design,crovetti2011finite}.
This allows an attacker to use high-frequency electromagnetic signals to bypass the differential signaling and further corrupt the transmitted information.

Note that for most protocols the transmitted information is binary (1 or 0), and the primary impact of the electromagnetic signals is to cause the receiver to detect bits incorrectly~\cite{sabath2010classification}.  
A few random manipulated bits can result in bit errors and disrupt the communications, but a sequence of manipulated bits can form an arbitrary malicious message, allowing the attacker to dictate the victim device's actions.
For example, imagine a scenario where an attacker continuously radiates electromagnetic signals to interfere with a wired USB keyboard in an office.
Such an attack can make a computer never receive valid messages from the keyboard because of bit errors.
Further, the attacker may carefully tune her attacking signals and radiate them at the proper time so as to manipulate bits in an organized sequence, and as such, she may inject an arbitrary command into the computer through the USB keyboard.

Realizing such an attack on differential signaling is not as easy as it sounds. It can be broken down into two main challenges: first, how to inject electromagnetic interference into the cable, and second, how to use the injected signal to make the victim device detect the intended bit.
Solutions to the first challenge have been studied in previous work, and we will introduce them in Section~\ref{sec:background}.
The second challenge is more subtle and depends on the target system. However differential signaling receivers work in largely the same way, independent from the application scenario or communication protocol, which makes it possible to use a single attack strategy to attack a large number of well-known communication protocols and applications. 

It needs to be pointed out that some studies already demonstrated electromagnetic signal injection attacks into a single-ended communication line~\cite{selvaraj2018intentional, selvaraj2018electromagnetic,dayanikli2021electromagnetic}, but the bit injection into differential signaling is a different process as we will show in this work.

We summarize our contributions as follows:
\begin{itemize}
\item We abstract and parameterize a generalized system model from practical circuits so as to capture the characteristics of different systems by tuning the parameters. We also define an adversary model, clarifying an attacker's objectives and capabilities that are essential to achieve attacks, as well as her limitations in practice. (Section~\ref{sec:model})
\item We detail the principles of how an electromagnetic signal can bypass the differential signaling technique and we further explain how the bypassed signal can arbitrarily manipulate bits that are recognized by the victim system. (Section~\ref{sec:fundamentals})
\item We quantitatively analyze the success rate of injections. We also discuss critical factors that need to be paid attention to in order to achieve a high success rate. (Section~\ref{sec:analysis_success_rate}).
\item We demonstrate the attacks on different chips experimentally to verify the attack principles (Section~\ref{sec:experiments}). Moreover, we successfully demonstrate how to use electromagnetic signal injection attacks to inject an arbitrary message into a CAN bus at a distance (Section~\ref{sec:can}).
\end{itemize}

The rest of this paper is organized as follows.
In Section~\ref{sec:background}, background on electromagnetic signal injection attacks is introduced, and in Section~\ref{sec:discussion}, a discussion is presented. 
We summarize related work in Section~\ref{sec:related_work}, and at last, a conclusion is drawn in Section~\ref{sec:conclusion}. 

\section{Background on Electromagnetic Signal Injection}
\label{sec:background}

As mentioned previously, wires/traces that are for signal transmission between/within circuits can also act as antennas to capture external electromagnetic waves.
The capture process is rather complicated, but it has been well studied in the area of ``Electromagnetism'', and further fully developed in ``Antenna Theory''.
Going into details about the capture process is beyond the scope of this work; however, in simple terms, the electromagnetic waves impact the metal conductors by inducing voltage changes in them. 
In this way, the electromagnetic waves are converted into electrical signals, further being superimposed with original electrical signals inside the wires and causing undue waveform changes.
Therefore, the antenna-like behavior of the wires allows an attacker to inject malicious signals into a circuit remotely.

Many researchers have thoroughly studied such electromagnetic signal injections and successfully demonstrated them in various systems~\cite{kasmi2015iemi,kune2013ghost,
rasmussen2009proximity,selvaraj2018electromagnetic,
Markettos2009Tfia, osuka2018information, giechaskiel2019framework, shin2016sampling, tu2019trick, sp20Zhang,wang2022ghosttouch, dayanikli2020senact, dayanikli2021electromagnetic,ware2017effects,selvaraj2018intentional, zhang2022detection, wang2022ghosttalk}.
From the previous studies, in order to achieve effective injections, two points need to be paid attention to.
First, it requires strong enough attack power.
There are many reasons that the attacking signals will be attenuated.
For example, the attacking signals will be attenuated during their propagations from the attacker's emitter to the victim system~\cite{balanis2015antenna, kune2013ghost}; when the attacking signals arrive at the victim system, RF shielding materials and filtering techniques that are deployed to protect the victim system from external interference will further attenuate the attacking signals.
As a result, the injected signal may be too weak to cause effective impacts on the victim system, and hence, enough attack power is essential.
Second, since the attack power cannot be infinite, it is also essential to consider how to maximize the injected power into the targeted wire.
Usually, this requires that the frequency of the attacking signal must be at the resonant frequency of the wire.
The resonant frequency of the wire can be approximated by its length; nonetheless, a better way to determine the resonant frequency is to have a copy of the receiving circuits and sweep through a range of frequencies~\cite{kune2013ghost}.
Note that since the lengths of the wires in the victim systems usually range from as short as several millimeters to meters long, the frequencies of the attacking signals are typically in the MHz and GHz frequency bands.

Since the previous studies already demonstrated that it is not hard to inject a malicious signal into a victim system by electromagnetic waves, we do not further detail the injection procedures hereafter.
However, in the remaining of this work, we will focus on explaining how the victim system responds to the injected signals and how an attacker can exploit the responses to inject arbitrary information into the victim system.

\section{System Model and Adversary Model}
\label{sec:model}

We abstract and parameterize a system model of differential signaling from practical circuits.
This model allows us to capture the characteristics of different circuits by tuning the parameters.
Next, we define an adversary model, which explains an attacker's capabilities and limitations.

\begin{figure}[t]  
  \centering
  \includegraphics[width=0.48\textwidth]{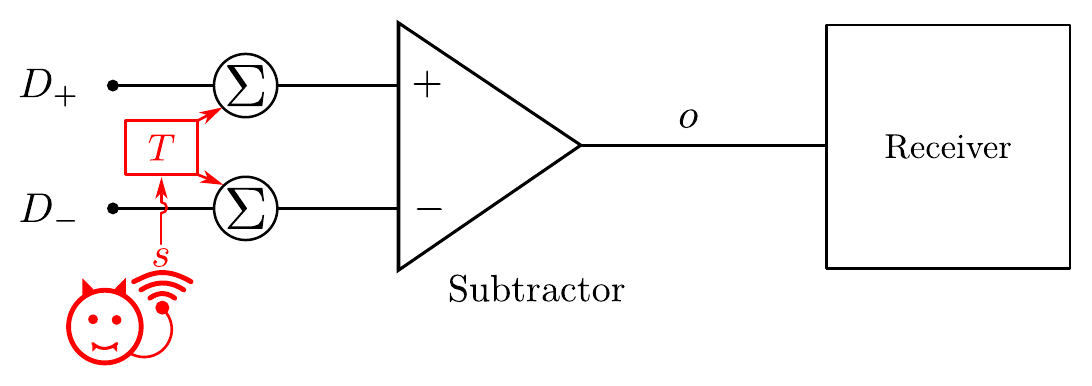}
  \caption{The system model consists of a pair of complementary signals ($D_{+}$ and $D_{-}$), a subtractor, and a receiver. The difference between the signals is extracted by the subtractor, and next, the difference is sent to the receiver. When an attack happens, the injected signals are superimposed with the complementary signals.}
  \label{fig:system_model}
\end{figure}

\subsection{System Model}
\label{sec:system_model}


Recall that information is carried by the difference between the differential pair of signals.
To obtain the difference, circuits that can ``subtract'' one signal from the other are used.
We define the circuits with such a subtraction function as a ``Subtractor'',  and a block diagram of the subtractor is shown in Figure~\ref{fig:system_model}. 
The subtractor has two inputs, each receiving a signal of the differential pair. 
The subtractor calculates the difference and then sends it to the following circuits for further processing.

After receiving the subtractor output signal, an essential step is to convert its analog voltage levels into a sequence of bits.
This is because the circuits that process information are usually digital (e.g., microprocessors) that only handle logic~1 and logic~0.
After obtaining the bits, other functions or tasks such as decoding, error checking, authentication, etc., can be further executed.
We define a ``Receiver'', as presented in Figure~\ref{fig:system_model}, to incorporate all functions of the circuits that handle the subtractor's output signal.

\subsubsection{Parameterizing Subtractor}
\label{sec:param_subtractor}

\begin{figure}[t]
     \centering
     \begin{subfigure}[b]{0.19\textwidth}
         \centering
         \includegraphics[width=\textwidth]{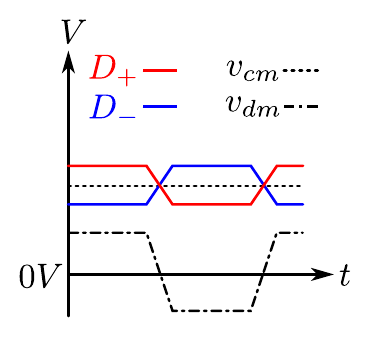}
         \caption{Subtractor}
         \label{fig:system_model_subtractor_input}
     \end{subfigure}
     \hspace{0.03\textwidth}
     \begin{subfigure}[b]{0.19\textwidth}
         \centering
         \includegraphics[width=\textwidth]{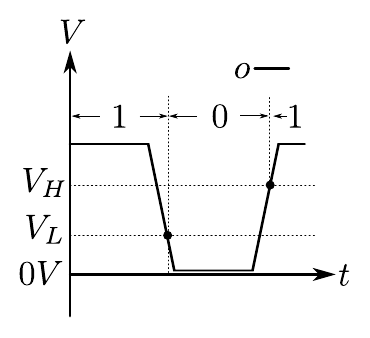}
         \caption{Receiver}
         \label{fig:system_model_receiver_input}
     \end{subfigure}
      \caption{(a) The subtractor's two input signals $D_{+}$ and $D_{-}$ can be represented by their differential mode $v_{dm}$ and common mode $v_{cm}$. (b) The receiver compares $o$ with two thresholds to determine the logic levels.}
      \label{fig:system_model_input}
\end{figure}

We denote the subtractor's two input signals as $D_{+}(t)$ and $D_{-}(t)$, and the output signal as $o(t)$.
To simplify the notations, we omit time $t$ hereafter.
In order to explicitly show the information that is carried by the two input signals, these two input signals can also be rewritten by their differential mode and common mode.
The differential mode is defined as the difference between two signals, and we denote it as ${v_{dm} = D_{+} - D_{-}}$, and it is $v_{dm}$ that represents the transmitted information; the common mode is defined as the average of two signals, and it is denoted as ${v_{cm} = \frac{D_{+} + D_{-}}{2}}$.
Such a relationship between the two input signals and their modes is illustrated in Figure~\ref{fig:system_model_subtractor_input}.
Note that $v_{cm}$ is a non-zero constant in almost all protocols, and thus, we assume that it is non-zero hereafter unless stated otherwise.

In practice, the subtractor is essentially a differential amplifier, and it is reasonable and prevalent to model its output signal as a sum of the amplified differential mode and the amplified common mode~\cite{razavi2002design}.
We denote the gains for the differential mode and the common mode as $G_{dm}$ and $G_{cm}$, respectively; note that the gains are functions of frequencies.
The amplified terms are expressed as ${G_{dm} \cdot v_{dm} + G_{cm} \cdot v_{cm}}$.

In addition, there also exist distortion and noise that contaminate the output signal.
The distortion originates from nonlinear properties of electronic components (e.g., transistors) that make up the subtractor~\cite{razavi2002design}, and we define a function ${F(v_{dm}, v_{cm})}$ to incorporate the impacts of the distortion phenomenon on the input signals.
We model the noise as additive Gaussian white noise, denoting it as $n$.
Thus, the subtractor output is expressed as:
\begin{equation}
\label{eq:diff_amp_output}
o  = G_{dm} \cdot v_{dm} + G_{cm} \cdot v_{cm} + F(v_{dm}, v_{cm}) + n 
\end{equation}

In Equation~\ref{eq:diff_amp_output}, the first term $G_{dm} \cdot v_{dm}$ explicitly carries the transmitted information, i.e., $v_{dm}$.
It needs to be emphasized that every subtractor has a finite operating frequency range, within which it is designed to function properly.
Inside this operating frequency range, the differential-mode gain remains constant, and as such the subtractor can guarantee a consistent output while handling input signals at different bit rates.
The common-mode gain is so small that it makes typical attenuation of $\SI{70}{\dB} - \SI{120}{\dB}$ to the common mode of the inputs~\cite{devices2009op}, thus making ${G_{cm} \cdot v_{dm}}$ nearly zero.
The distortion of the subtractor is also well maintained, and thus the impact of $F(v_{dm}, v_{cm})$ is negligible.
Therefore, the subtractor is sufficiently good enough at rejecting the impacts of the common mode within the operating frequency range. 
However, this no longer holds out of the operating frequency range, and we will detail the reasons in Section~\ref{sec:fundamentals}.

\subsubsection{Parameterizing Receiver}
\label{sec:param_receiver}

The primary function of the receiver is to convert analog voltages into bits as mentioned previously.
It is a common way in practical circuits that the logic levels are determined by comparing analog voltage levels with two pre-determined thresholds~\cite{harris2010digital}.
The reason for using two thresholds instead of a single one is that the difference between two thresholds can prevent the noise from causing wrongly detected logic levels.
We denote these two thresholds as $V_{H}$ and $V_{L}$, and ${V_{H} > V_{L}}$.
The detection rule is straightforward: as depicted in Figure~\ref{fig:system_model_receiver_input}, when ${o \geq V_{H}}$, a logic 1 will be detected; when ${o \leq V_{L}}$, a logic 0 will be detected. Specifically, when $o$ is between these two thresholds (e.g., the noise causes the voltages to fluctuate into this region), the detected bit will retain its value.
Note that the receiver detects bits periodically.

In addition, it is also essential to cover the circuits before the logic level detection, as an attacker needs to exploit these circuits to achieve a wrongly detected bit. 
We will detail the attack principles in Section~\ref{sec:impact_receiver}, but here, we abstract a model and explain the functions of the circuits.

When a signal enters the receiver, it first goes through an electrostatic discharge (ESD) circuit, which is commonly used to protect the input pins of any electronic device from overvoltages.
A block diagram of the ESD circuit is presented in Figure~\ref{fig:charge_discharge_transistor}: it clamps the negative overvoltages to a minimum allowed voltage (e.g., ${GND}$), and the positive overvoltages to a maximum allowed voltage (e.g., $V_{DD}$)~\cite{redoute2009emc}. 
After that, a buffer circuit follows. It is used to get rid of an impedance mismatch between the previous stage and the receiver, and more precisely, it provides isolation and prevents undesired interaction from the previous stage~\cite{carter2001handbook}.
The buffer circuit is essentially built up with transistors, which work like switches, and its function is abstracted in a way as illustrated in Figure~\ref{fig:charge_discharge_transistor}: its input signal controls the switches, generating an output signal to reproduce its input signal.
In this way, the buffer circuit transfers its input signal to the logic level detection.
In all, the circuits before the logic level detection are modeled as a combination of the ESD circuit and the buffer circuit.

\subsection{Adversary Model}
\label{sec:adversary_model}

An attacker's objective is to inject a message with a length of $L$ bits into a victim system.
The attacker has no physical access to the victim system, implying that she cannot modify its circuits, nor can she tap wires to inject attacking signals into it.
Because of no physical access, it is rather difficult to know which bit is transmitted in the wires.
However, we do not limit the attacker's ability to guess the bit, and we will further explain and discuss her guess ability in detail in Section~\ref{sec:analysis_success_rate}. 
The attacker can wirelessly inject the attacking signals into the victim system by radiating electromagnetic waves, and she can tune her attacking signals, regarding their frequencies, power, etc.
The attacker knows the period that the receiver detects a bit.
In each bit injection, the duration of the attacking signal is set the same as the period, and hence, the attack can always interfere with the receiver when it detects a bit.

As mentioned previously, an electromagnetic signal injection is a complicated process in practice. 
We define a transfer function $T$ to explain any changes to the attacking signal $s(t)$ due to the injection process, e.g., frequency selectivity, attenuation, spreading, etc.
To simplify the notation, we omit time $t$, and thus the injected signal is denoted as $T(s)$.
Note that there could exist multiple injection places in the victim system.
However, only when the injection impacts the signals that carry the information will the attacker be able to manipulate the bits, which are recognized by the receiver. 
Therefore, it is equivalent to modeling the pair of wires as the injection point as shown in Figure~\ref{fig:system_model}.
In fact, the differential signaling technique is usually deployed between two ends that are far from each other, and the pair of wires are effective antennas capturing the attacking signals in practice.

\section{Bit Injection Attack}
\label{sec:fundamentals}

\begin{figure}[t]
  \centering
  \includegraphics[width=0.48\textwidth]{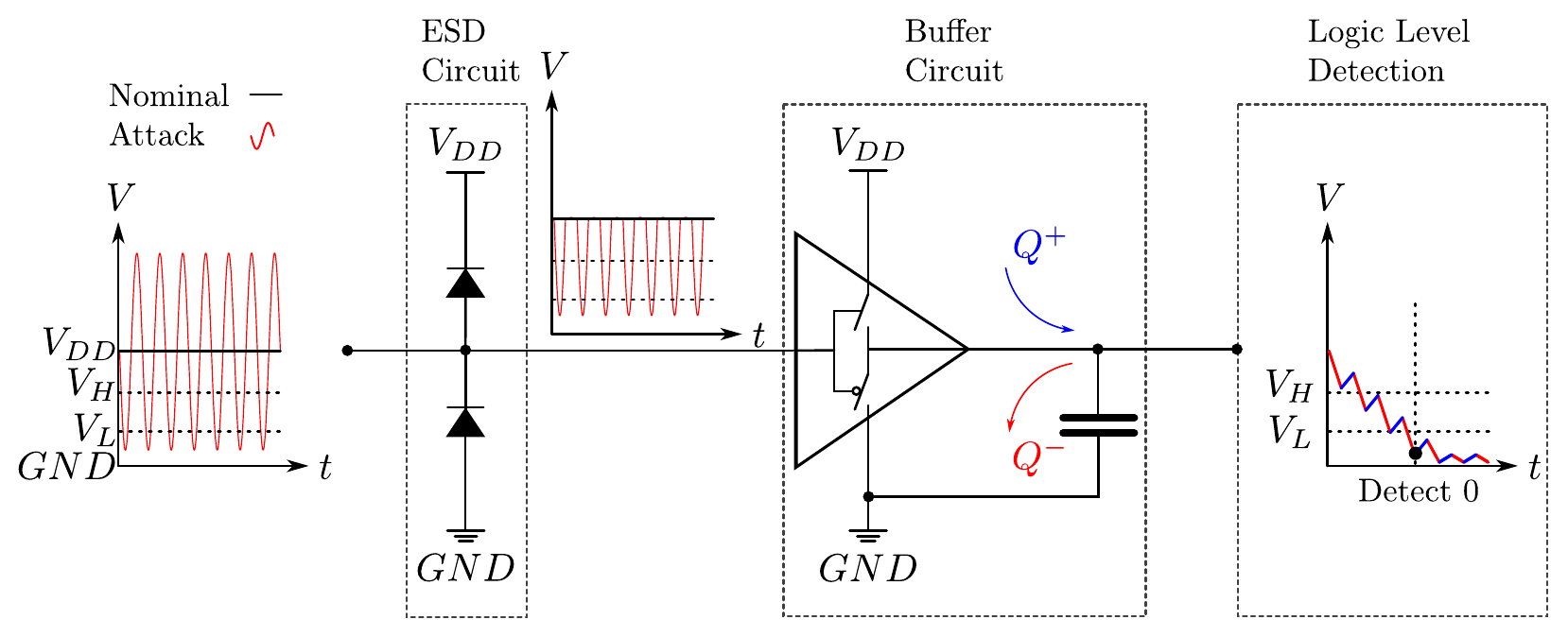}
  \caption{This diagram shows a model of circuits before logic level detection in the receiver. The charging and the discharging of the output parasitic capacitance are asymmetrical, and a net charge~${Q^{+} - Q^{-}}$ causes the output voltage changes when an attack happens.}
  \label{fig:charge_discharge_transistor}
\end{figure}

In this section, we will detail how an attack can make the victim system detect an incorrect bit.
Note that as mentioned in Section~\ref{sec:background}, since previous research has thoroughly studied the signal injection process, we do not further detail it here.
However, we will focus on explaining why injected signals can bypass the subtractor, and next, how the receiver responds to the bypassed injected signals.

\subsection{Bypassing Subtractor}
\label{sec:bypass_subtractor}

The injected signals are superimposed with the two input signals of the subtractor, leading to ${D_{+} + T(s)}$ and  ${D_{-} + T(s)}$.
According to the definition of the differential mode and the common mode in Section~\ref{sec:param_subtractor}, the differential mode will not be affected because the injected signals cancel each other out; however, an extra term $T(s)$ that is the average of the identical injected signals is added to the common mode.
As explained previously, the common mode of the original input signals has almost no impact on the subtractor output signal. 
Thus, under the interference of the attack, it is equivalent to approximate the common mode to be $T(s)$, or ${v_{cm} = T(s)}$.
Substitute it into Equation~\ref{eq:diff_amp_output}, and we can obtain an expression of a malicious subtractor output signal $o'$ as follows:
\begin{equation}
\label{eq:diff_amp_output_attack}
o'  = G_{dm} \cdot v_{dm} + \\ G_{cm} \cdot T(s) + F(v_{dm}, T(s)) + n 
\end{equation}
Note that the terms ${G_{cm} \cdot T(s) + F(v_{dm}, T(s))}$ are malicious changes that are caused by the injected signal $T(s)$, and these terms explicitly represent the bypassed injected signal. 

It is essential to point out that the subtractor's common-mode rejection ability is finite, and it is ascribed to two widely accepted reasons.
First, the two inputs of the subtractor are not perfectly symmetrical in practice. 
This results in that common-mode variations in the inputs are converted to differential-mode variations in the output, which is also known as ``common-mode to differential-mode conversion''~\cite{meyer1966common,jaeger1976common,yi1980common,giustolisi2000cmrr,razavi2002design}.
Second, nonlinearities of the subtractor lead to extra unexpected variations in the output~\cite{crovetti2006finite,crovetti2011finite}.
Especially beyond the operating frequency range of the subtractor, on the one hand, the subtractor's common-mode rejection ability deteriorates dramatically because the common-mode to differential-mode conversion becomes more and more significant at higher frequencies~\cite{razavi2002design}, and this indicates that $G_{cm}\cdot T(s)$ becomes larger.
On the other hand, the nonlinear phenomenon of the subtractor becomes stronger~\cite{crovetti2011finite}, leading to much more distortion, or larger $F(v_{dm}, T(s))$.
The evidence above indicates that if the injected signals are out of the operating frequency range,  their impacts on the common mode are much more easily converted into additionally malicious voltage changes in the subtractor's output.
In this way, the injected signals bypass the subtractor.

It is essential for the attacker to know the waveform of the bypassed injected signal so that she can have a controllable impact on the next stage, i.e., the receiver.
However, since the subtractor is not initially designed for use beyond the operating frequency range, it is not easy to precisely predict the waveform.
To tackle this challenge, a determined attacker can get a replica of the subtractor and experimentally find the relationship between the output signal and the input signals of the subtractor, and as such, the attacker can estimate the bypassed injected signal.
We will demonstrate it in Section~\ref{sec:exp_subtractor}.

\subsection{Bit Detected Incorrectly in Receiver}
\label{sec:impact_receiver}

Recalling in Section~\ref{sec:param_receiver}, we explained that the receiver determines a bit by comparing the subtractor output signal with two thresholds. 
In order to make the receiver detect an incorrect bit, the bypassed injected signal must make the nominal voltage level (which is represented by $G_{dm} \cdot v_{dm}$) of the subtractor's output signal cross the threshold that determines the opposite bit.
Since detecting an incorrect bit is literally flipping a bit, we will use these two synonyms interchangeably hereafter.
To make the explanation concise, we assume that the nominal voltage is at $V_{DD}$, which is above $V_{H}$, and it means that the receiver is supposed to receive 1 if no attack presents.

The oscillation of the bypassed injected signal either pushes the voltage level towards or away from $V_{L}$, as shown in Figure~\ref{fig:charge_discharge_transistor}.
However, only when the malicious voltage change causes the voltage level to move towards $V_{L}$ will the receiver wrongly detect a bit.
Luckily, the ESD circuit guarantees the direction of the malicious voltage change.
This is because the positive part of the bypassed injected signal exceeds the maximum allowed voltage, leading to rectification by the ESD circuit, but the negative part remains, and thus the voltage level moves toward $V_{L}$, as shown in the middle of Figure~\ref{fig:charge_discharge_transistor},. 

After being rectified, the malicious voltage change continues propagating through the buffer circuit of the receiver. 
Note that the frequency of such a malicious signal is further beyond the operating frequency range of the receiver.
The fast oscillation of the malicious voltage change will make the switches close and open periodically, thus charging and discharging the parasitic capacitance of its output periodically. 
However, the charging process and the discharging process are asymmetrical, leading to a quick accumulation of net charge across the output capacitance, and then, it holds still~\cite{fiori2011susceptibility, crovetti2013ic}.
If a bit is read when the voltage level crosses $V_{L}$, 0 is recognized.

In this way, the bypassed injected signal successfully makes the receiver detect 0.
In a similar vein, when the nominal voltage level is below $V_{L}$ (and the receiver expects 0), the bypassed injected signal also makes the receiver detect 1.
Researchers pointed out that the impacts of the malicious voltage change are equivalent to adding a constant DC offset to the input signal of the receiver~\cite{fiori2011susceptibility, crovetti2013ic}.
The magnitude of this equivalent DC offset depends on the frequency and amplitude of the malicious voltage change, as well as the specific circuits that are impacted~\cite{fiori2011susceptibility}.
This implies that an attacker can successfully flip a bit by properly choosing the frequency and the power of her attacking signals.
However, it is also not easy to predict the receiving circuits' responses out of their operating frequency ranges, and hence, it will be difficult to figure out a formula to calculate the effective frequency and power of the attacking signals.
Still, a determined attacker can experimentally obtain such attacking signals by sweeping the frequency and power to find ranges of effective attacking signals.
We will demonstrate this in Section~\ref{sec:exp_receiver}.

\section{Analysis of Success Rate}
\label{sec:analysis_success_rate}

After knowing the principles of a bit injection attack, we then analyze its success rate.
When an attacker intends to inject a bit, it is essential to consider which bit is being transmitted in the wire:
if the transmitted bit is not what the attacker wants, she needs to emit an attacking signal so as to flip the bit; otherwise, she does not need to emit any attacking signal, leaving the bit unchanged.
However, it is not an easy job to know which bit is transmitted in practice, and we will further discuss some methods in Section~\ref{sec:discussion}.
Still, the attacker can make use of her knowledge about the victim system to make a guess of the bit, and her guess will further dictate her actions.
To make a successful injection, on the one hand, it is essential to have a correct guess; on the other hand, the attacking signal can effectively flip a bit.
In this section, we first parameterize the attacker's guess, as well as the effectiveness of her attacking signals, and next, we will analyze the success rate.

\subsection{Parameterization}
\label{sec:parameterization}
We denote the bit that is transmitted in the wire as $A$, and the attacker's guess as $G$.
We use a parameter $g$ to quantify the attacker's knowledge about~$A$, and ${g \in \left[0, 1\right]}$.
We define that~${g = \frac{1}{2}}$ means the attacker knows nothing about the bit, and there is an equivalent chance that the attacker will guess~1~or~0.
Furthermore, we define that~${g > \frac{1}{2}}$ means the attacker knows information that indicates the bit could be 1.
A larger~$g$ means that the attacker knows more information, and thus, it is more likely to guess 1; when $g = 1$, the attacker is sure that the bit is 1.
Conversely, ${g < \frac{1}{2}}$ means the attacker knows information that indicates the bit could be 0, and a smaller $g$ also implies knowing more information, and thus, it is more likely to guess 0; when $g = 0$, it means the attacker is sure that the bit is 0.
We model that $G$ follows a Bernoulli distribution with the parameter $g$, where $G$ takes 1 with a probability of $g$ and 0 with a probability of $1 - g$.

We quantify the performance of an attacking signal by two parameters: $u$ represents the probability of flipping 1 to 0, and $v$ represents the probability of flipping 0 to 1, and $u, v \in \left[0, 1\right]$.
For a certain victim system, each attacking signal corresponds to a pair of $u$ and $v$, and all $u, v$ pairs together characterize this specific victim system's responses to attacks.
We define \emph{feasible pairs} to incorporate all these pairs.
In practice, $u$ and $v$ can be measured experimentally, and we will demonstrate the measurements and the characterization in Section~\ref{sec:exp_receiver}.
In addition, here are two special pairs that need to be paid attention to.
The first pair is $u = 0$ and $v = 1$. 
Since $1 - u = 1$ means that a logic 1 is always kept unchanged and $v = 1$ means that a logic 0 will always be flipped successfully, the corresponding attacking signal will force any bit to 1.
Conversely, a pair of $u = 1$ and $v = 0$ corresponds to an attacking signal that can force any bit to 0.
With these two ideal pairs, the attacker can inject any bit successfully without any guess all the time.
Unfortunately, they are not always attainable in practice and we will show it in Section~\ref{sec:exp_receiver}.

\subsection{Success Rate of Bit Injection}
\label{sec:success_rate_bit}
Let's begin by considering that the attacker intends to inject a single 1.
There are four combinations of $A$ and $G$, and the attacker's actions and the success rate for each combination are as follows:
\begin{itemize}
\item If $A = 1$ and $ G = 1$, the attacker makes a correct guess, and since she intends to inject 1, she will not radiate any attacking signal. The success rate is 1, which can be written as~$A \cdot G$.
\item If $A =0$ and $G = 1$, the attacker wrongly thinks that the bit is already 1 so that she will not radiate any attacking signal, meaning that she will never flip the bit. Hence, the success rate is~0.
\item If $A = 1$ and $G = 0$, the attacker wrongly thinks that the bit is 0 and she will radiate an attacking signal. However, the attacking signal needs to keep the bit unchanged such that it is still 1. Thus, the success rate is ${1 - u}$, which can be written as~${A \cdot (1 - G) \cdot (1 - u)}$.
\item If $A = 0$ and $G = 0$, the attacker's guess is correct, and the attacker will radiate an attacking signal to flip the bit. The success rate of flipping 0 is $v$, which can also be written as~${(1-A) \cdot (1 - G) \cdot v}$.
\end{itemize}
We denote the success rate of injecting 1 as $P_{1}$, and it can be expressed as a combination of these cases:
\begin{equation*}
\label{eq:p1}
P_{1} =
\begin{cases}
\mbox{$G + (1 - G) \cdot (1 - u)$,} & \mbox{if $A = 1$} \\
\mbox{$(1 - G) \cdot v$,} & \mbox{if $A = 0$} \\
\end{cases}
\end{equation*}

Suppose the attacker intends to inject a single 0, we can use a similar way to reach an expression of the success rate of injecting 0, which is denoted as $P_{0}$ and expressed as:
\begin{equation*}
\label{eq:p0}
P_{0} =
\begin{cases}
\mbox{$G \cdot u$,} & \mbox{if $A = 1$} \\
\mbox{$(1 - G) + G \cdot (1 - v)$,} & \mbox{if $A = 0$} \\
\end{cases}
\end{equation*}
We will focus on the injection of 1 hereafter, as the injection of 0 is a symmetrical process and the explanation is similar.

\subsubsection{Impact of $g$}
\label{sec:success_rate_g}

To investigate the impact of $g$, we start from the expectation of $P_{1}$, which can be easily derived and expressed~as:
\begin{equation*}
E(P_{1}) =
\begin{cases}
\mbox{$u \cdot g + 1 - u$,} & \mbox{if $A = 1$} \\ 
\mbox{$- v \cdot g + v$,} & \mbox{if $A = 0$} \\ 
\end{cases}
\end{equation*}
Essentially, the larger $E(P_{1})$ is, the better.
Since we are discussing the impact of $g$, it is reasonable to assume that $u$ and $v$ are non-zero here; otherwise, $g$ vanishes in $E(P_{1})$.

If~$A = 1$, $E(P_{1})$ increases with~$g$.
According to our definition of $g$, a bigger $g$ means knowing more information about~$A = 1$, and thus it is more possible to make a correct guess.
The importance of making a correct guess can be easily proved: if $A = 1$, $P_{1}$ is maximized when ${G = A}$.
Thus, it can be concluded that if~$A = 1$, the larger $g$ is, the more possible that $P_{1}$ will be maximized, and the better.
Similarly, if~$A = 0$, $E(P_{1})$ increases while decreasing~$g$, and a smaller~$g$ means a higher chance of making a correct guess, and thus more possible to maximize~$P_{1}$.

Regarding $P_{0}$, the analysis is similar and we do not further detail it here.
Therefore, to make a correct guess to maximize the success rate, two points need to pay attention to: first, it is crucial that $g$ is in a manner conforming with $A$, and second, it is always better to know more information about $A$.

\subsubsection{Impact of $u$ and $v$}
\label{sec:success_rate_uv}

As indicated by the equation of $P_{1}$, the larger ${1-u}$ and $v$ are, the better.
However, it needs to be emphasized that in a specific system, $u$~and~$v$ are related in a certain way, and an example is shown in Figure~\ref{fig:microbit_uv_pairs}.
From our experiments with different chips in Section~\ref{sec:exp_receiver}, we observe that there is a trade-off between increasing~${1-u}$ and increasing~$v$.
Then, here comes the question: Which is the optimal pair?

Determining the optimal pair can be formulated into a multi-objective optimization problem, where $1-u$ and $v$ are the objectives.
The most extensively used method of solving such an optimization problem is called the weighted sum method~\cite{marler2010weighted,branke2008multiobjective}, where the two objectives are combined and converted into one scalar, composite objective function by assigning proper weights to them; note that the sum of the weights equals 1.
We select $g$ as the weight for $1 - u$, and $1 - g$ as the weight for $v$, and the reasons are as follows.

Firstly, if the attacker has no knowledge about $A$ (where~${g = \frac{1}{2}}$), it is equivalently important to ``keep 1 unchanged'' and ``flip 0''.
Hence, it requires that the weights are equal, and~${g = 1-g = \frac{1}{2}}$ meets the requirement.
Secondly, if the attacker knows information indicating that the bit is 1 (where~${g >\frac{1}{2}}$), ``keeping 1 unchanged'' is more important, and hence, more weight for ${1-u}$ than $v$.
Moreover, when more information is known, the importance of $1 - u$ further increases, and so does the weight. 
Since ${g > 1 - g}$ and knowing more information also means that $g$ increases, $g$ can properly quantify the weight of ${1-u}$, and accordingly, ${1 - g}$ quantifies the weight of $v$.
Thirdly, if the attacker knows information indicating that the bit is 0, we can also deduce that $g$ as the weight for ${1-u}$ and ${1- g}$ as the weight for $v$ in a similar way, and we do not further detail the reason. 
Therefore, searching for the optimal pair of $u$~and~$v$ of injecting 1 is solving the following problem:
\begin{align*}
\max_{(u, v)} \quad & g \cdot (1 - u) + (1 - g) \cdot v\\
\textrm{s.t.} \quad & (u, v) \in \textrm{feasible pairs}
\end{align*}
In a similar vein, concerning injecting 0, the larger $u$ and $1-v$ are, the better.
Finding the optimal $u$ and $v$ of injecting 0 is solving the following problem:
\begin{align*}
\max_{(u, v)} \quad & g \cdot u + (1 - g) \cdot (1 - v)\\
\textrm{s.t.} \quad & (u, v) \in \textrm{feasible pairs}
\end{align*}

In Section~\ref{sec:exp_receiver}, we will demonstrate how to use the method above to find the optimal pairs and then verify that the optimal pairs will do better than other pairs.
Note that since the attacker has no access to the victim system, when she is preparing attacking signals, she needs to conduct experiments on a replica and use the methods above to find the optimal pairs.

\subsubsection{Measuring Susceptibility}
Although the attacker cannot access the victim system, a system designer of the victim system can do so.
Thus, she can measure and obtain the optimal pairs, and then, use them to estimate the success rate. 
Note that the success rate is also a metric that sufficiently quantifies the susceptibility of the victim system: a higher success rate means that the victim system is more susceptible to an injection; conversely, a lower success rate means less susceptible.
Thus, the system designers can use this analysis to quantitatively evaluate the security of their systems, and are able to change components or data modulation scheme to reduce adversarial success.

\subsection{Success Rate of Message Injection}
\label{sec:success_rate_message}

Recall that the attacker’s objective is injecting $L$ bits into the victim system, and the success rate of injecting a message will decreases exponentially with the message length.
However, it needs to be pointed out that in Section~\ref{sec:impact_receiver}, we explained that an attack can cause voltage changes to accumulate quickly and then holds still.
Therefore, suppose the attacker will perform identical attacks (i.e., the same attacking signal, the same intended injected bit) on a sequence of transmitted bits that are consecutive and identical, once the first bit injection is successful, the success rates of the following bit injections will increase.
This is because the first successful bit injection attack gets rid of many uncertainties with respect to the guess, the effectiveness of the attacking signal, timing, etc.
We can approximate the success rate after the first successful bit injection to be 1 until the end of the consecutive injections.
Such an approximation may overestimate the success rate of a message injection, as unpredictable responses in the victim system may still lead to a failure of a bit injection.
An example of using such a method to estimate the success rate of a message injection will be shown in Section~\ref{sec:can}.
It needs to be emphasized that system designers would rather overestimate the success rate than underestimate it because when they deploy measures to improve the security the nominal protection will make the victim system less susceptible in practice.

\section{Experiments}
\label{sec:experiments}

Recalling in Section~\ref{sec:fundamentals}, we explain the principles of injecting bits. 
In this section, we will demonstrate that injected signals can easily bypass the subtractor. 
Then, we will show that the bypassed injected signals cause incorrectly detected bits in the receiver.

\subsection{Testbed}
\label{sec:exp_testbed}
\begin{figure}[t]
  \centering
  \includegraphics[width=0.38\textwidth]{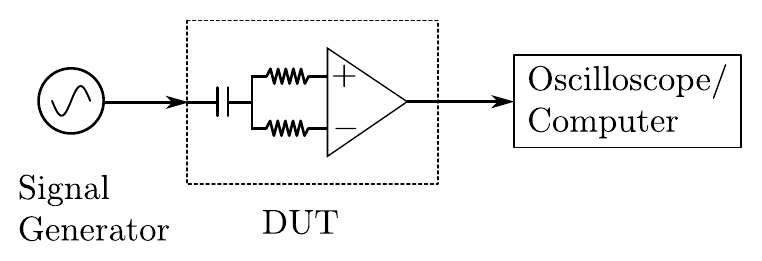}
  \caption{A testbed consists of a signal generator, a device under test (DUT), and an oscilloscope/computer.}
  \label{fig:experiment_setup}
\end{figure}

To test different chips of the subtractor and the receiver, we first build a testbed before our experiments. 
The testbed's functions are generating attacking signals and measuring responses of a device under test (DUT).

A setup of the testbed is shown in Figure~\ref{fig:experiment_setup}.
A signal generator produces an attacking signal and injects it into the DUT through a wire.
Such a signal injection setup is also known as Direct Power Injection (DPI) methodology~\cite{giechaskiel2019sok}, and we use it because the injected frequency and power can be precisely controlled so that we can measure the responses of the chips reliably.
Moreover, we use an oscilloscope to capture and measure the waveforms of the injected signals and the DUT's output signals, and a computer is used to process and analyze the measurements.

\subsection{Subtractor}
\label{sec:exp_subtractor}

We choose five different off-the-shelf subtractor chips,  which are TJA1050, MCP2551, SN65HVD230, MX485, and SN751768P. 
They support CAN bus or RS485/422, and they are widely used in many critical applications such as automotive, medical equipment, and industrial devices.
The subtractor chip is configured in a way as shown in the DUT block in Figure~\ref{fig:experiment_setup}: two same resistors are added to the input of the subtractor, and these two transistors are equivalent to the terminated resistors in practice that are required by the differential signaling standards.
Note that the voltage difference between the two inputs is internally configured to keep unchanged.

Regarding the injected signal, it is coupled to the midpoint between the two resistors by a capacitor.
Doing so is equivalent to injecting a common-mode interference into the subtractor.
The injected signal is sinusoidal, and its frequency is swept from $\SI{10}{\kilo\hertz}$ to $\SI{100}{\mega\hertz}$, and its peak-to-peak voltage of the injected signal is set to be $\SI{1}{\volt}$, $\SI{2}{\volt}$, and $\SI{4}{\volt}$.
Note that other waveforms such as square and sawtooth are potentially effective, but due to the limit of our signal generator, we cannot produce high-frequency and powerful signals with these special waveforms, so we stick to sinusoidal signals in our experiments.

\subsubsection{Impacts of Injected Frequency and Power}
\label{sec:exp_subtractor_power_frequency}

\begin{figure}[t]
	\centering
    \includegraphics[width=0.48\textwidth]{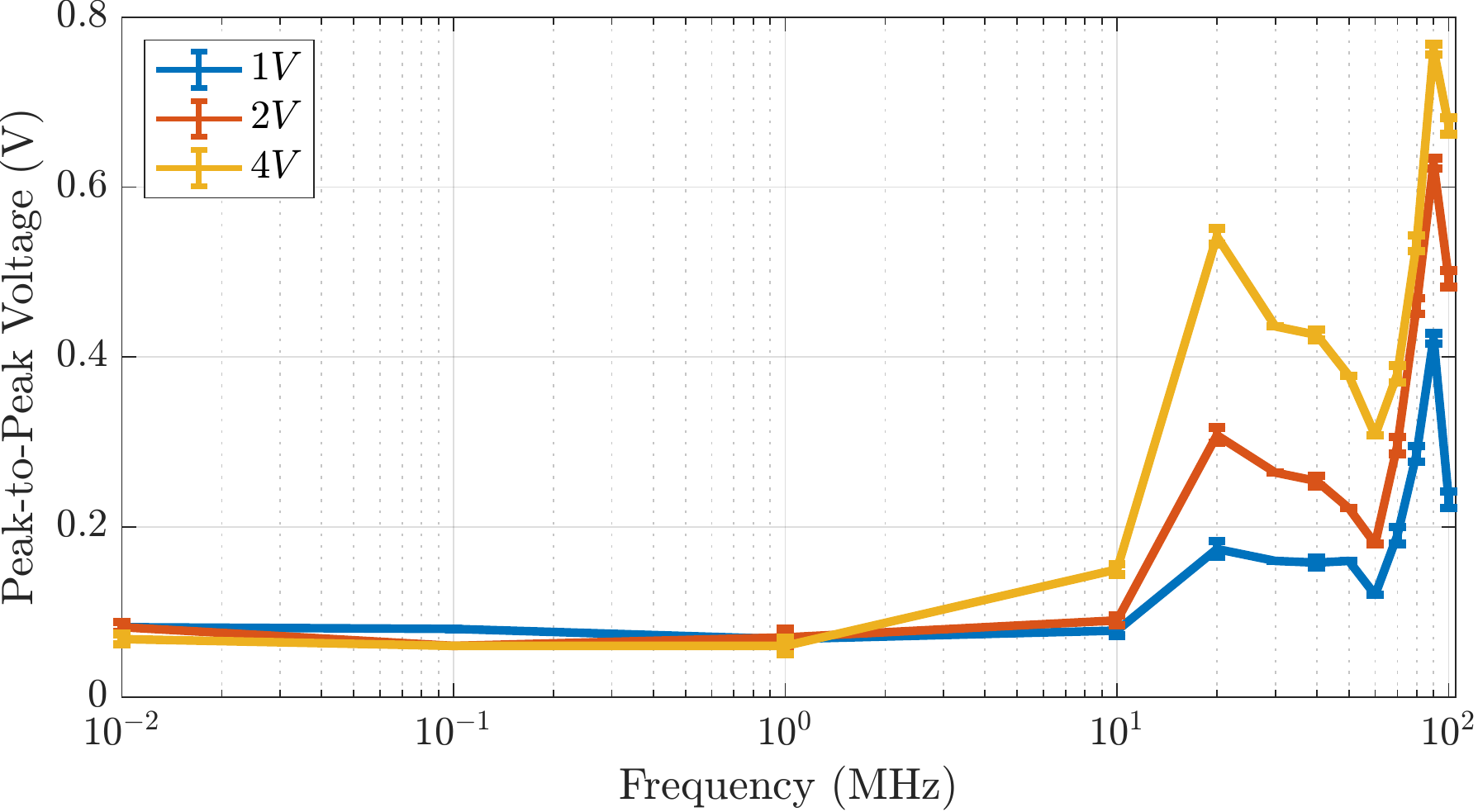}
    \caption{Test a subtractor chip TJA1050 with injected signals with frequencies ranging from $\SI{10}{\kilo\hertz}$ to $\SI{100}{\mega\hertz}$ and peak-to-peak voltages from $\SI{1}{\volt}$ to $\SI{4}{\volt}$. The y-axis represents the output of the subtractor.}
    \label{fig:tja1050}
\end{figure}

As explained in Section~\ref{sec:param_subtractor}, when no attack happens, the subtractor's output signal remains consistent with the differential mode of its input signals.
In our configuration above, since the voltage difference between the two inputs is constant, the subtractor's output signal is also constant.
Note that the noise essentially exists, but it is too small to significantly disturb the output signal. 
When an injected signal applies, the subtractor's output signal will start oscillating, and such an oscillation represents the bypassed injected signal.
Note that the bypassed injected signal is explained by ${G_{cm} \cdot T(s) + F(v_{dm}, T(s))}$ in Equation~\ref{eq:diff_amp_output_attack}.
Therefore, we can use the peak-to-peak voltage of the subtractor's output signal to quantify the strength of the bypassed injected signal.

Taking a subtractor chip TJA1050 as an example, when there is no attack, the peak-to-peak voltage of the output signal is $\SI{0.06}{\volt}$, which reflects the noise level.
When an attacking signal is injected into the chip, the averaged peak-to-peak voltage and its standard deviation are shown in Figure~\ref{fig:tja1050}.
Between $\SI{10}{\kilo\hertz}$ and $\SI{1}{\mega\hertz}$, the output is as close as the noise level.
This is because the common-mode interference is well handled within the operating frequency range.
However, when the frequency is increased above $\SI{10}{\mega\hertz}$, the peak-to-peak voltage has an increasing trend along with the frequency.
These results explicitly show that the subtractor's common-mode rejection ability deteriorates out of the operating frequency range.
In addition, two local maximums appear at $\SI{20}{\mega\hertz}$ and $\SI{90}{\mega\hertz}$, as shown in Figure~\ref{fig:tja1050}.
This means that the injected signals at these two frequencies bypass this subtractor chip more efficiently than other frequencies.
From the perspective of an attacker, she can take advantage of properly choosing the injected frequency to achieve a bypass using less attack power.

\begin{figure}[t]
	\centering
    \includegraphics[width=0.48\textwidth]{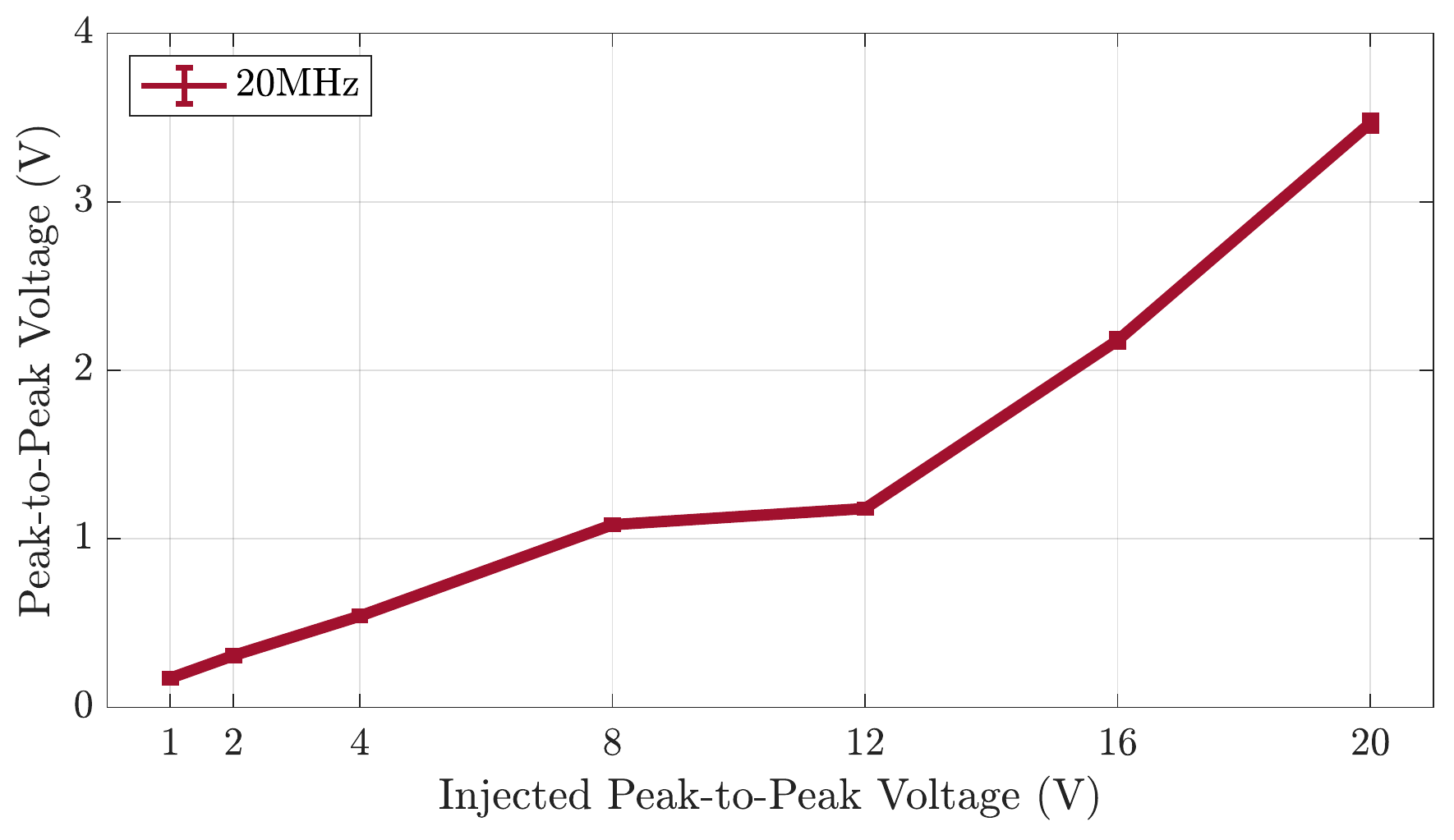}
    \caption{The strength of the bypassed injected signal increases while the injected power. }
    \label{fig:tja1050_20mhz}
\end{figure}

While increasing the injected power, the peak-to-peak voltage of the output also increases, implying a stronger bypassed injected signal.
However, as shown in Figure~\ref{fig:tja1050}, with the injected power of $\SI{4}{\volt}$, the highest peak-to-peak voltage of the bypassed injected signal is still below $\SI{1}{\volt}$.
To achieve a stronger bypassed injected signal in the subtractor output, we use an RF power amplifier to increase the injected signal up to $\SI{20}{\volt}$ at $\SI{20}{\mega\hertz}$, which is an efficient frequency that the subtractor lets the injected signal bypass.
The output of the subtractor is shown in Figure~\ref{fig:tja1050_20mhz}.
The results indicate that with increasing the injected power, the strength of the bypassed injected signal also increases.
Also, it can be observed that the strength of the bypassed injected signal is roughly proportional to the injected power, and this allows the attacker to estimate the strength of the bypassed injected signal.

Note that such a bypassing phenomenon does not only occur in the TJA1050 chip but also in many other subtractor chips.
For example, in the other four chips we tested, it is observed that the injected signal can always bypass them when the frequency is increased out of their operating frequency ranges; also, they all show that the higher the injected frequency/power is, the stronger the bypassed injected signal is.

\subsubsection{Impacts of Noise and Distortion}
\label{sec:exp_subtractor_distortion}

\begin{figure}[t]
	\centering
    \includegraphics[width=0.48\textwidth]{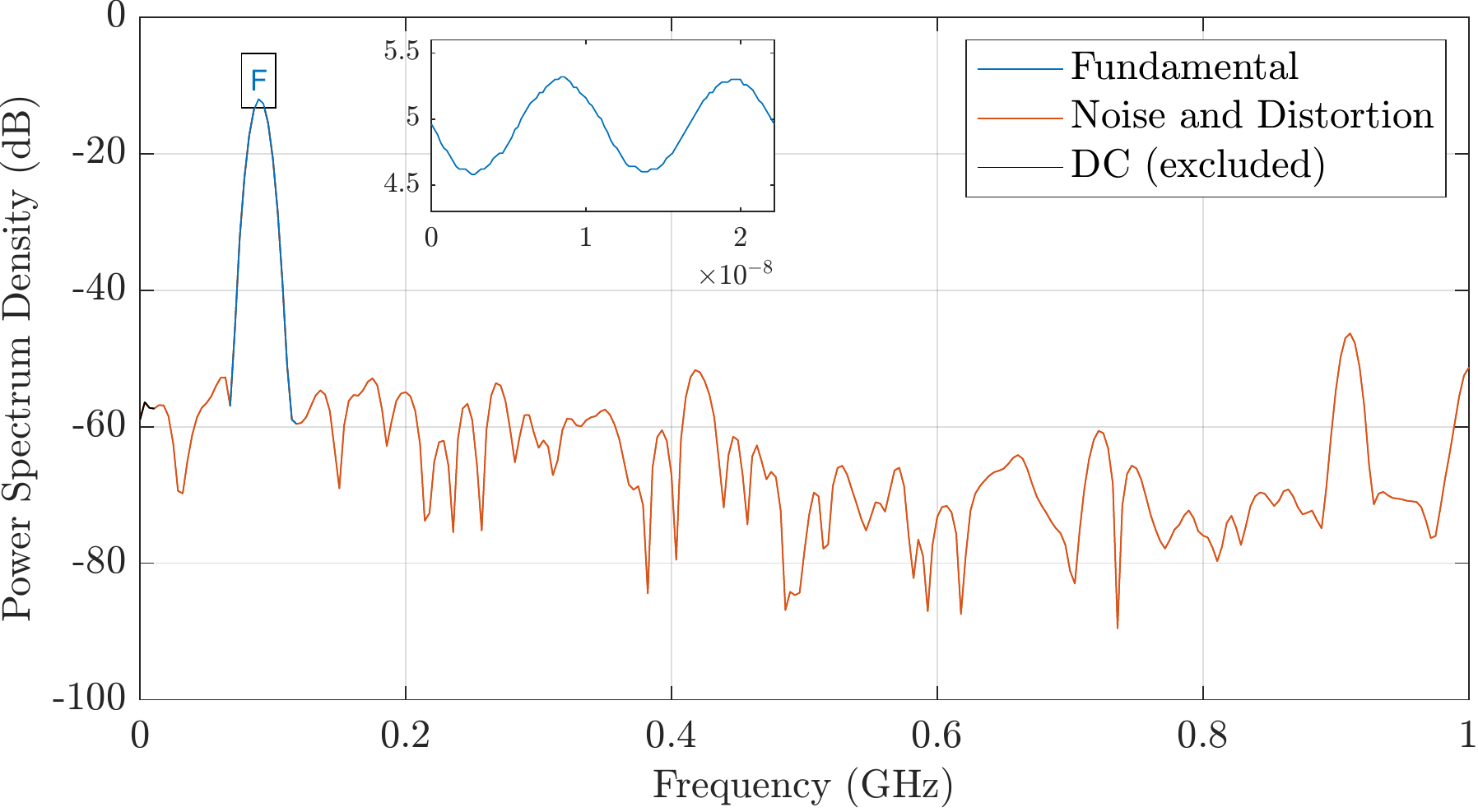}
    \caption{Set the injected frequency to be $\SI{90}{\mega\hertz}$ and the injected power to be $\SI{4}{\volt}$. The frequency domain of the bypassed injected signal is presented. A time domain screenshot of the bypassed signal is in the floating window, where the x-axis is time (s) and the y-axis is voltage (V).}
    \label{fig:tja1050_fdomain}
\end{figure}

As indicated by Equation~\ref{eq:diff_amp_output_attack}, the distortion ${F(v_{dm}, T(s))}$ plus the noise $n$ will make the bypassed injected signal differs from the injected signal regarding waveforms.
It is essential to know how much the bypassed injected signal is distorted because the bypassed injected signal will act on the receiver straight away and its waveform determines how the receiver responds.
 
To measure the impacts, we use a signal to noise-and-distortion (\emph{SINAD}) ratio as a metric, which is calculated by the following equation:
\begin{equation}
SINAD = 10\times\log_{10}\frac{P_{s}}{P_{n+d}}
\end{equation}
where $P_{s}$ is the power of the fundamental frequency of the signal, and $P_{n+d}$ is the power of noise and distortion.
The \emph{SINAD} ratio is a widely used measure that quantifies the quality of a signal that is particularly degraded by the noise and distortion~\cite{zumbahlen2008linear,kester2009understand}.
The higher the \emph{SINAD} ratio of a signal is, the better the signal quality is, and hence, less distortion in the signal.
In Figure~\ref{fig:tja1050_fdomain}, a frequency domain of a bypassed injected signal is presented to show the difference between the fundamental frequency and the noise plus the distortion, and the \emph{SINAD} is around $\SI{27}{\dB}$.
In this figure, the distortion exists in the bypassed injected signal as harmonics, but they are too small to distort the bypassed injected signal significantly, which can also be observed from the time domain of the signal (please refer to a floating window at the top-left corner in the figure).

\begin{figure}[t]
	\centering
    \includegraphics[width=0.48\textwidth]{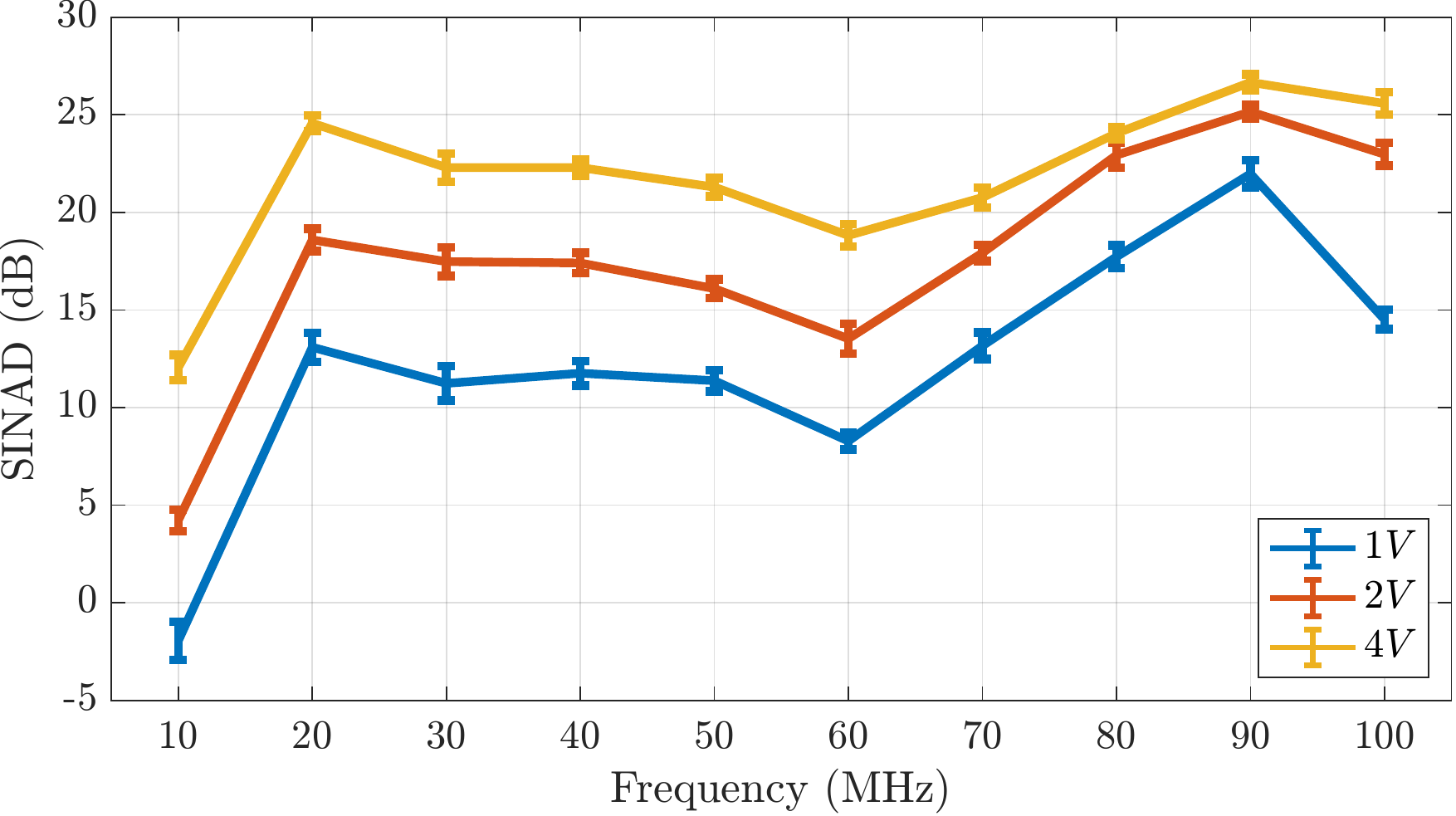}
    \caption{The signal to noise-and-distortion (\emph{SINAD}) ratio quantifies the signal quality of the bypassed injected signal. Specifically, it measures how much the fundamental frequency component of the bypassed injected signal is stronger than the noise plus distortion. The larger the ratio, the less the signal is distorted.}
    \label{fig:tja1050_sinad}
\end{figure}

We also take the TJA1050 chip as an example, and in Figure~\ref{fig:tja1050_sinad}, we show its \emph{SINAD} ratio when different injected signals apply.
The ratio is low when the injected frequency and the injected power are small (e.g., $\SI{10}{\mega\hertz}$ and $\SI{1}{\volt}$), and this is because only a tiny amount of injected signal can bypass the subtractor, as explained previously.
While either increasing the injected power or the frequency, the ratio has an increasing trend.
In addition, the \emph{SINAD} ratio is at least $\SI{10}{\dB}$ for most of the measurements.
Such a result implies that this chip demonstrates weak distortion and noise, which do not need to be worried too much while modeling its output signal.
However, it does not mean that every subtractor chip has such weak distortion and noise, and an attacker still needs to handle them carefully.

\subsection{Receiver}
\label{sec:exp_receiver}

\begin{figure}[t]
	\centering
    \includegraphics[width=0.40\textwidth]{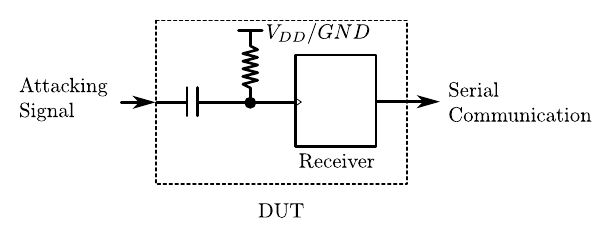}
    \caption{Change the DUT for a receiver: a pull-up (or -down) resistor is used to fix the DC level of the digital input pin at a certain level ($V_{DD}$ or $GND$), and the receiver sends measurements to a computer by serial communication.}
    \label{fig:experiment_setup_receiver}
\end{figure}

\begin{figure}[t]
     \centering
     \begin{subfigure}[b]{0.23\textwidth}
         \centering
         \includegraphics[width=\textwidth]{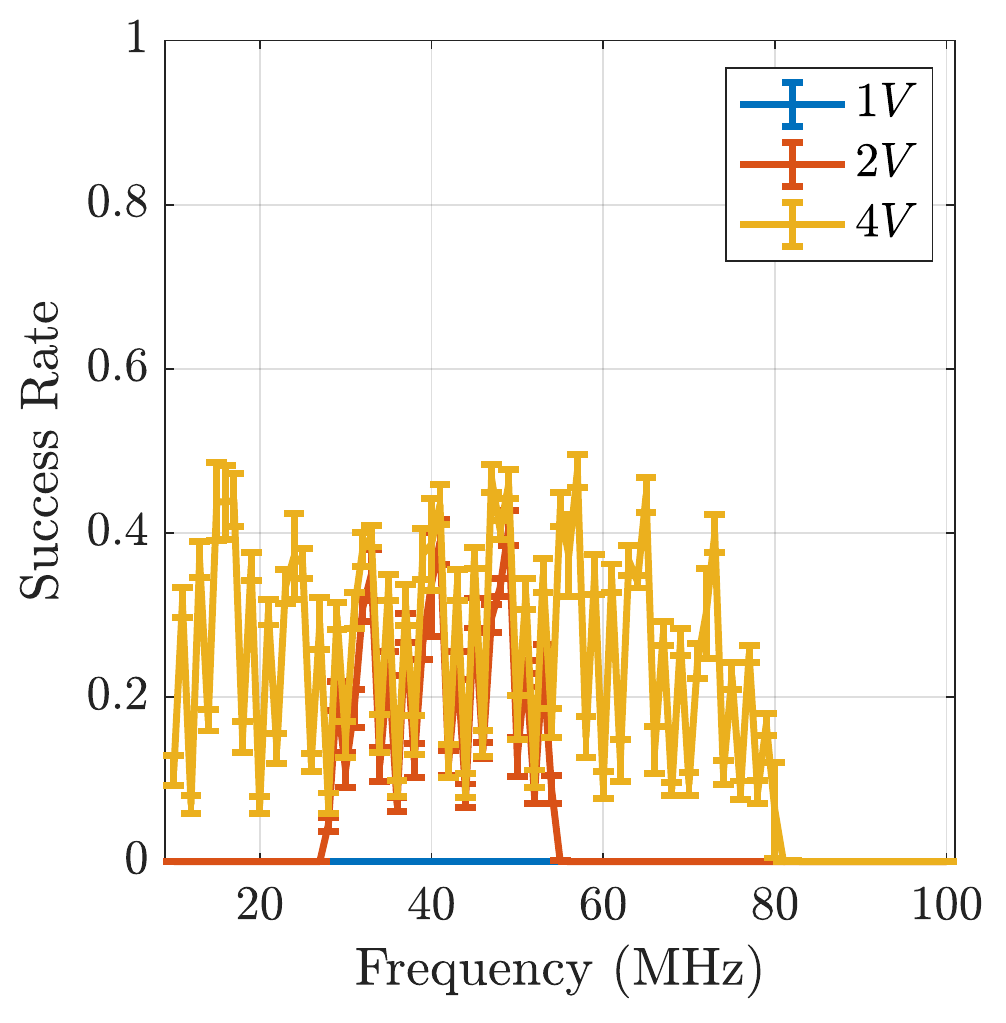}
         \caption{}
         \label{fig:microbit1to0}
     \end{subfigure}
     \begin{subfigure}[b]{0.23\textwidth}
         \centering
         \includegraphics[width=\textwidth]{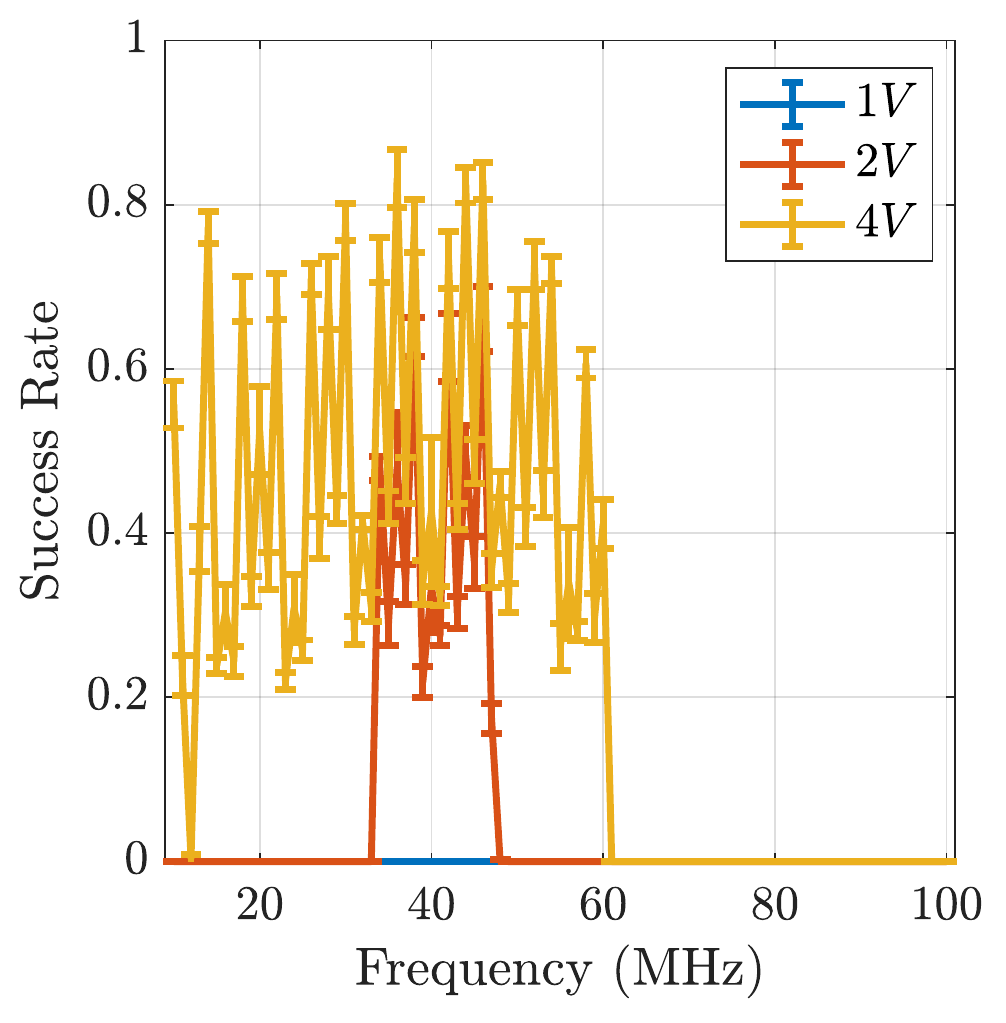}
         \caption{}
         \label{fig:microbit0to1}
     \end{subfigure}
      \caption{Success rates of bit injections in nRF52833. (a) Flip bits from 1 to 0. (b) Flip bits from 0 to 1.}
      \label{fig:microbit}
\end{figure}

In various systems, microcontrollers are usually the devices that realize the receiver functions: they detect the logical level of the input signals and then execute specific tasks according to the received information.
We select three different microcontroller chips to test, which are nRF52833, ATMEGA328P, and ATSAM3X8E.
We use the testbed that is shown in Figure~\ref{fig:experiment_setup} to study how the injected signal impacts bits that are recognized by the receiver chips.
However, there are small modifications in the DUT block, and they are shown in Figure~\ref{fig:experiment_setup_receiver}.
First, the input is changed to single-ended. 
Second, a pull-up (or -down) resistor is used to fix the input voltage level at a high (or low) voltage level, which corresponds to logic 1 (or 0).
Third, because these chips support serial communication with the computer, the DUT directly sent recognized bits to the computer through a serial communication line.

We set the injected frequency from $\SI{10}{\mega\hertz}$ to $\SI{100}{\mega\hertz}$ with a step of $\SI{1}{\mega\hertz}$.
Note that since the subtractor chips have demonstrated that they can well remove the common-mode interference below $\SI{10}{\mega\hertz}$, we do not further test that frequency range.
The peak-to-peak voltage of the injected signal is set to be $\SI{1}{\volt}$, $\SI{2}{\volt}$, and $\SI{4}{\volt}$. 
For each combination of the injected power and the injected frequency, 10 measurements are recorded;
in each measurement, 256 bits are collected by the chip, and we calculate the percentage of successfully flipped bits as the success rate; then, the mean and the standard deviation of the success rates are calculated and presented.

Taking an nRF52833 chip that works at $V_{DD} = \SI{3}{\volt}$ as an example, 
it has ${V_{H} = \SI{2.1}{\volt}}$ and ${V_{L} = \SI{0.9}{\volt}}$ according to its datasheet~\cite{nordic2009nrf}.
Recalling in Section~\ref{sec:param_receiver}, $V_{H}$ and $V_{L}$ are two thresholds that are used to determine logic levels.
To flip 1, the voltage change needs to be at least ${\SI{3}{\volt} -\SI{0.9}{\volt} = \SI{2.1}{\volt}}$; conversely, to flip 0, it needs to be at least ${\SI{2.1}{\volt} - \SI{0}{\volt} = \SI{2.1}{\volt}}$.
The experimental results of flipping 1 are shown in Figure~\ref{fig:microbit1to0}, and the results of flipping 0 are shown in Figure~\ref{fig:microbit0to1}.

When the injected signal is $\SI{1}{\volt}$, no bit flip is observed.
This is because the injected signal is too weak to cause the voltage change beyond the threshold.
When the injected signal is increased above $\SI{2}{\volt}$, bit flips happen.
Although the injected signal of $\SI{2}{\volt}$ is still weaker than the required threshold of $\SI{2.1}{\volt}$, recall that as explained in Section~\ref{sec:impact_receiver} the voltage change can accumulate quickly and lead to a voltage change over the threshold ultimately, and consequently, the bit flips happen.
When the injected power is increased to $\SI{4}{\volt}$, the success rate becomes higher.
Also, the frequency range where bit flip happens widens when the injected signal becomes much stronger.
The results also imply that this chip is more susceptible in a frequency range that is centered at $\SI{40}{\mega\hertz}$, and it is relatively easier to cause bit injections in this frequency range with less attack power. 

\begin{figure}[t]
	\centering
    \includegraphics[width=0.48\textwidth]{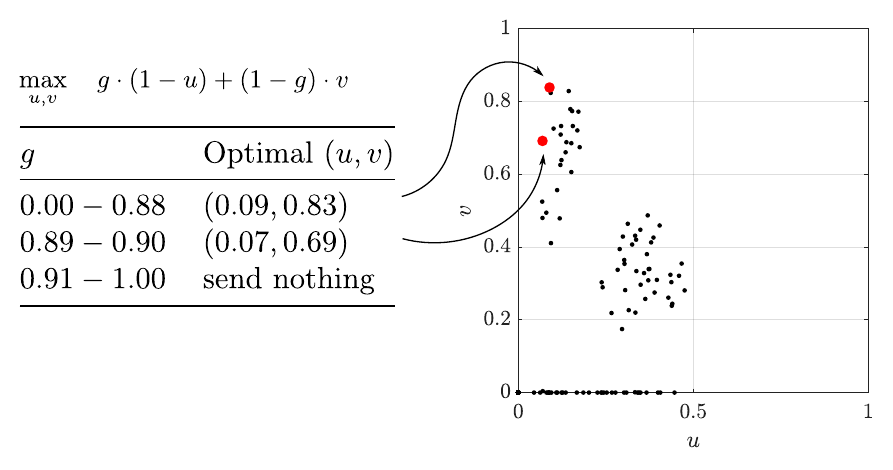}
    \caption{The pairs of $u$ and $v$ characterize the chip's responses to the attacks. Regarding injecting 1, the optimal pair with respect to $g$ can be decided by solving the optimization problem.}
    \label{fig:microbit_uv_pairs}
\end{figure}

In the other two chips, it is also observed that the success rates of bit injections are related to both the power and the frequency of the injected signal.
The results show that the higher the power is, the higher the success rate is, and the wider the frequency range in which bit flips happen.
Note that regarding the chip nRF52833 in Figure~\ref{fig:microbit}, the success rates show a periodic pattern in terms of the injected frequency: a peak appears every $\SI{2}{\mega\hertz}$.
Such a repeated pattern has nothing to do with the testing circuits outside the chip because the periodic pattern is not observed in other chips.
It is speculated that some deterministic properties of the nRF52833 chip lead to this periodic pattern.
However, it is trivial to figure out what these deterministic properties are because this periodic pattern only exists in this chip, and knowing the deterministic properties does not help attacks on other chips.

\subsubsection{Characterizing Receiver's Response}

Recalling in Section~\ref{sec:analysis_success_rate}, we introduce parameters $u$ and $v$, and they can be used to characterize a victim device's responses to attacks. 
It is not difficult to find that the success rate of flipping 1 is $u$ (see Figure~\ref{fig:microbit1to0}), and the success rate of flipping 0 is $v$ (see Figure~\ref{fig:microbit0to1}).
Thus, we can obtain the feasible pairs of $u$ and $v$, and we plot them in Figure~\ref{fig:microbit_uv_pairs}, which visualizes the chip's (nRF52833) responses to the attacks. 

As mentioned previously $u = 0$ and $v = 1$ are an ideal pair, which represents an attacking signal that forces any bit to 1.
The closer a pair is to it, the easier the injection of 1 will be.
Similarly, $u = 1$ and $v = 0$ is the other ideal pair, which represents an attacking signal that forces any bit to 0.
As shown in Figure~\ref{fig:microbit_uv_pairs}, the feasible pairs' distribution is skewed to $u = 0$ and $v = 1$, meaning that it is much easier to inject 1 than 0 into this chip.
Since injecting 1 and injecting 0 are symmetrical processes and the analysis will be similar, we focus on injecting 1 hereafter.

Recalling in Section~\ref{sec:success_rate_uv}, we formulate the method of determining the optimal pair of $u$ and $v$.
To determine the optimal pair regarding injecting 1, we assume that $g$ is always correct, and we present the results in Figure~\ref{fig:microbit_uv_pairs}.
When $g$ is below $0.88$, the optimal pair is $u = 0.09$ and $v = 0.83$.
Such an attacking signal can successfully flip 0 with a probability of $0.83$ and keep 1 unchanged with a probability of~${1 - 0.09 = 0.91}$.
With further increasing~$g$, as explained in Section~\ref{sec:success_rate_uv}, the attacker is becoming more and more sure that the bit is 1, and hence, $u$ decreases to $0.07$.
When ${g}$ is greater than 0.9, the solution indicates that the attacker will send nothing.
Next, we simulate attacks and show how the optimal pair outperforms.

\subsubsection{Simulation and Success Rate}
\begin{figure}[t]
     \centering
     \begin{subfigure}[b]{0.23\textwidth}
         \centering
         \includegraphics[width=\textwidth]{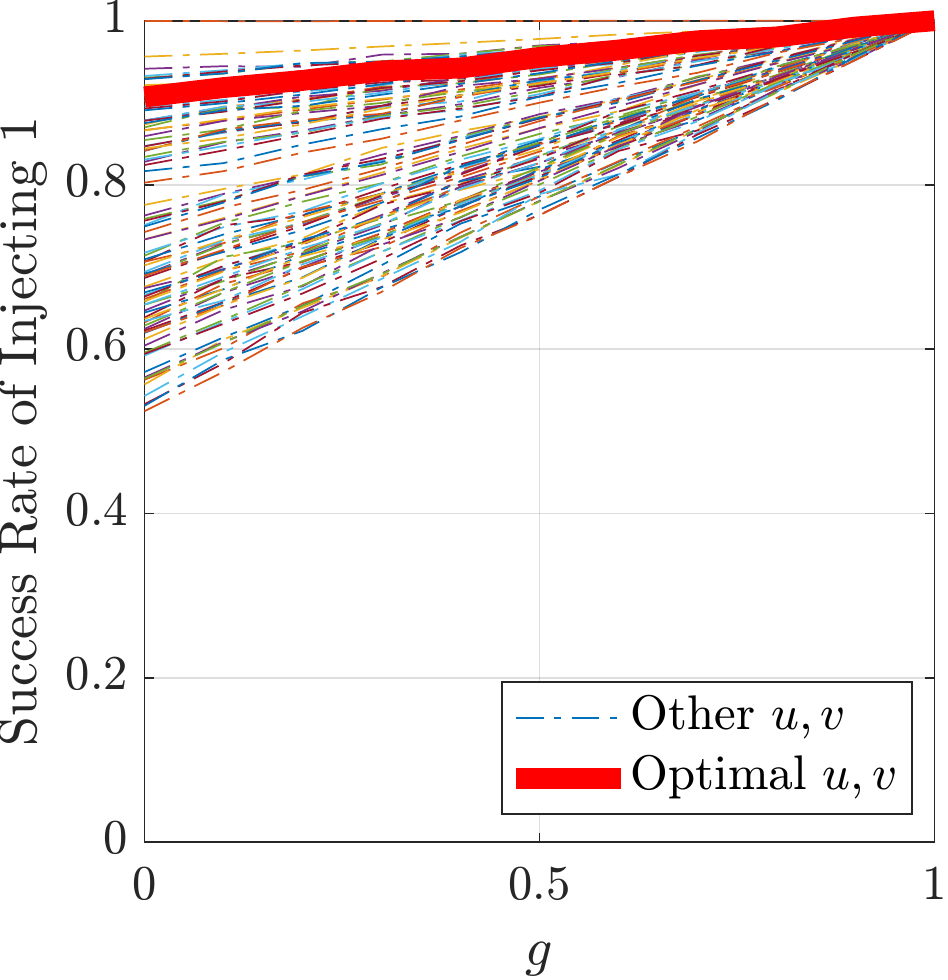}
         \caption{}
         \label{fig:microbit_successrate_inject1_a_1}
     \end{subfigure}
     \begin{subfigure}[b]{0.23\textwidth}
         \centering
         \includegraphics[width=\textwidth]{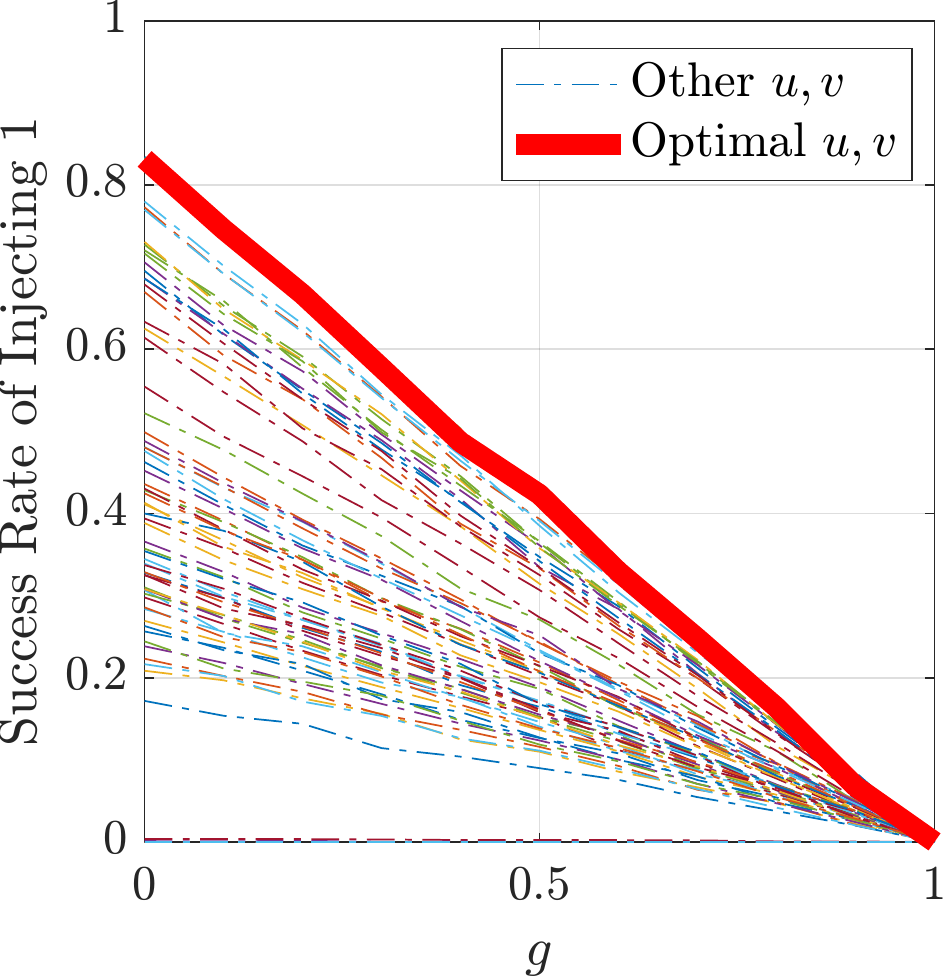}
         \caption{}
         \label{fig:microbit_successrate_inject1_a_0}
     \end{subfigure}
      \caption{(a) If $A = 1$, the success rate of injecting 1 with using different pairs of $u, v$. (b) If $A = 0$, the success rate of injecting 1 with using different pairs of $u, v$.}
      \label{fig:microbit_success}
\end{figure}

First, we simulate attacks with the optimal pair. 
The transmitted bit $A$ is set to either 1 or 0, and $g$ ranges from 0 to 1. 
We average the simulated success rates of each $g$ and present the results in Figure~\ref{fig:microbit_success}.
In Figure~\ref{fig:microbit_successrate_inject1_a_1}, when $A = 1$, the success rate increases with~$g$;
in Figure~\ref{fig:microbit_successrate_inject1_a_0}, when $A = 0$, the success rate decreases with~$g$.
The simulation results match with our model of $E(P_{1})$ in Section~\ref{sec:success_rate_g}, and in addition, the importance of having a~$g$ that is in a manner conforming with $A$ is also explicitly shown.

Next, we repeat the simulation with other pairs of $u$ and $v$, and compare them with the optimal pair, and the results are shown in Figure~\ref{fig:microbit_success}.
In Figure~\ref{fig:microbit_successrate_inject1_a_1}, when $A = 1$, some pairs outperform the optimal pairs, but these pairs are those that have small $u$ and small $v$: they are good at keeping 1 unchanged, but they cannot flip 0 effectively.
Therefore, as shown in Figure~\ref{fig:microbit_successrate_inject1_a_0}, when $A = 0$, the optimal pair outstrips others.

To decide whether the optimal pair outperforms any other pair significantly, we can conduct multiple t-tests.
Since the success rate has a linear relationship with $g$ as shown in both Figure~\ref{fig:microbit_successrate_inject1_a_1} and Figure~\ref{fig:microbit_successrate_inject1_a_0}, we use the averaged success rate at $g = \frac{1}{2}$ as a metric to represent the attack performance. 
Note that the simulation is repeated 100 times for each pair, thus 100 samples for each pair.
Next, we run t-tests to test against the alternative hypothesis that the optimal pair has a higher averaged success rate, or namely, outperforms the other pair.
The significance level is set to $0.05$, which is conventionally accepted as the threshold.
These tests show that they reject the null hypothesis, except the pair of $u = 0.092$ and $v = 0.82$.
It is not surprising because it is the pair that is close to the optimal pair of $u = 0.09$ and $v = 0.83$, as shown in Figure~\ref{fig:microbit_uv_pairs}.


\section{Message Injection into CAN}
\label{sec:can}

A Controller Area Network (CAN) is a protocol that is devised to allow many devices to communicate with each other on a two-wire bus, and it is now deployed in many different applications from medical instruments to automotive.
The CAN is a broadcast type of bus, and any device, also known as a node, can freely send/receive data.
This feature makes it possible that an attacker broadcasts whatever she wants on a CAN bus.
In this section, we first briefly introduce the basics of the CAN, and then we demonstrate how to inject an arbitrary message into the CAN.

\subsection{CAN Basics}

\begin{figure}[t]
	\centering
    \includegraphics[width=0.40\textwidth]{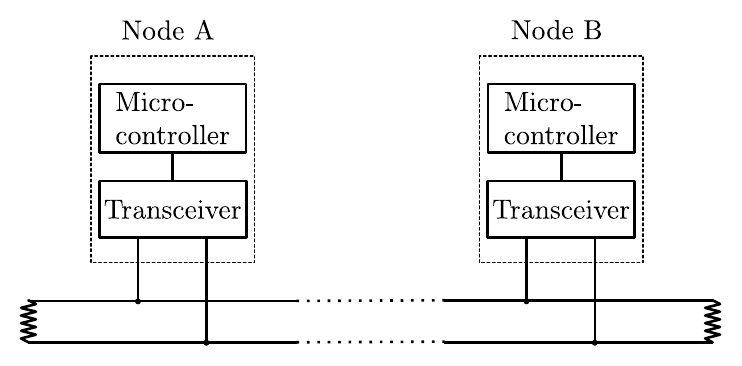}
    \caption{In the CAN system, nodes are connected to the same bus, where two wires are terminated by resistors.}
    \label{fig:can_system}
\end{figure}

A basic structure of the CAN is shown in Figure~\ref{fig:can_system}.
In a node, a transceiver is an interface between the wires and the microcontroller, and its function is to convert the differential signals into a single signal that the microcontroller can use while receiving data, or the other way around while transmitting data.
The microcontroller handles signals on a software level, including identifying the type of the data, error checks, bus arbitration, etc.

On the physical level, when the voltage levels of the differential signals are the same, a recessive state (1) is defined; otherwise, a dominant state (0).
Note that when no message is broadcast, the CAN system always remains at the recessive state.

\subsection{Message Injection}
\label{sec:message_injection}

\begin{figure}[t]
	\centering
    \includegraphics[width=0.48\textwidth]{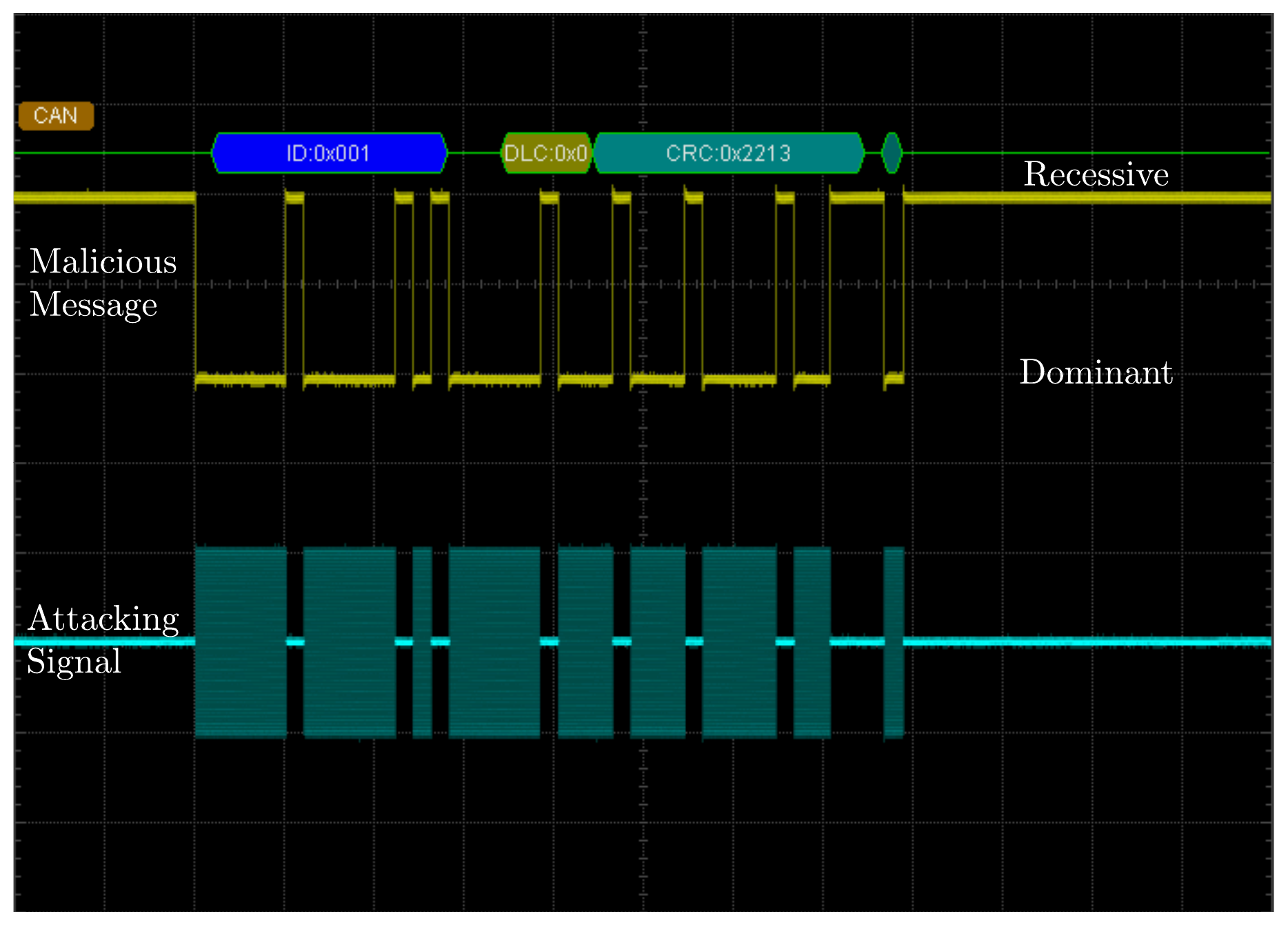}
    \caption{The attacker wants to inject a malicious message, and she generates an attacking signal according to the malicious message.}
    \label{fig:can_malicious_signal}
\end{figure}

\begin{figure}[t]
	\centering
    \includegraphics[width=0.48\textwidth]{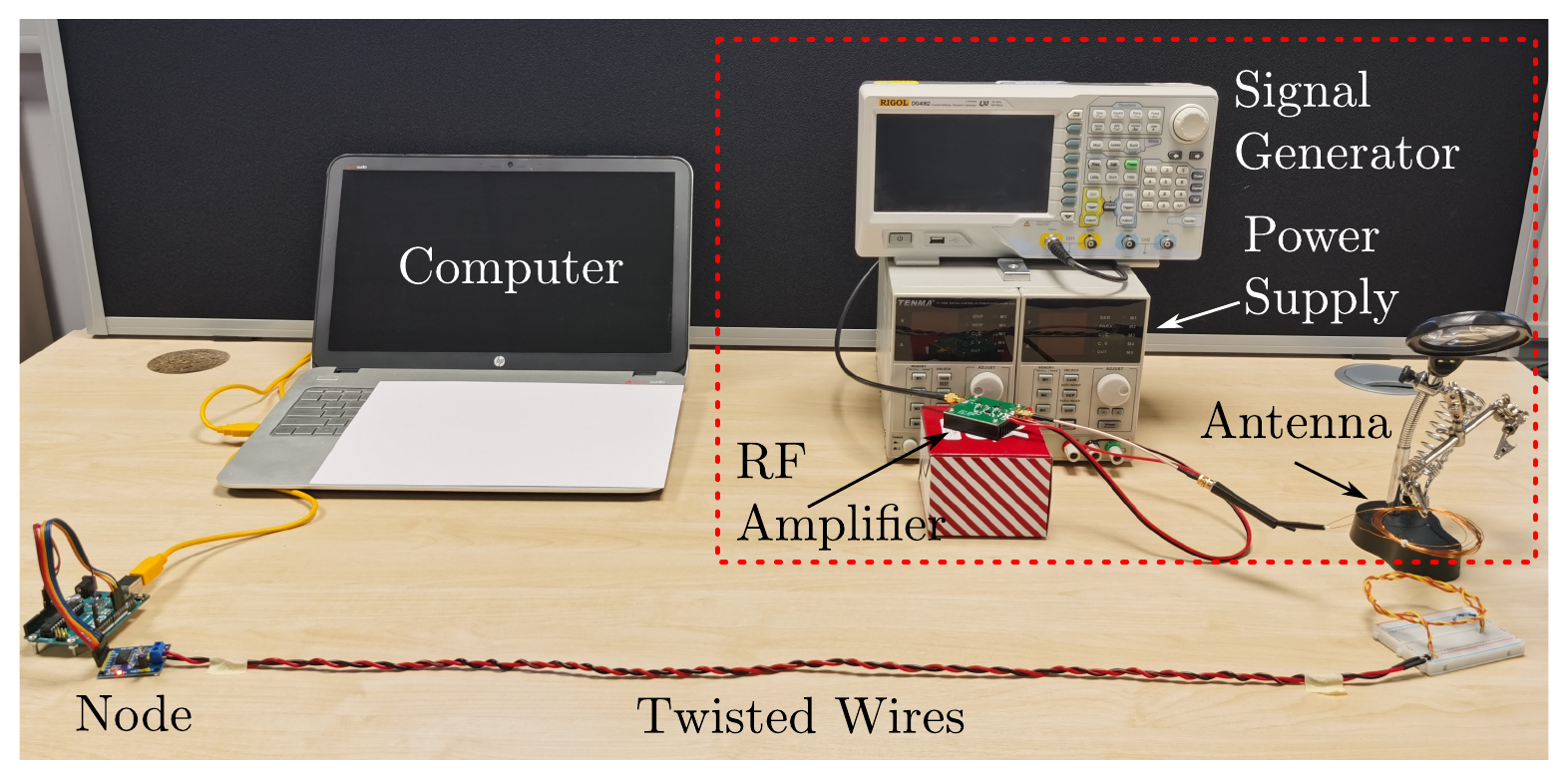}
    \caption{A practical setup of message injection attack on a CAN bus. The devices in the red rectangle form an attacker's setup}
    \label{fig:can_attack_real_setup}
\end{figure}

It is not difficult to find that such a CAN system matches our system model: the two wires in the CAN system correspond to the two input wires in our system model; the transceiver is the subtractor; the microcontroller is the receiver.
Thus, it is possible for an attacker to use the bit injection attack to inject arbitrary messages into the CAN system.
We detail the attack as follows.

We assume that the attacker has $g = 1$, i.e., she knows the line is always at a recessive state.
With an arbitrary message that the attacker wants to inject, the first step is to convert it into a sequence of bits according to the rules of the CAN protocol.
Based on the bits, an attacking signal is generated.
For example, the attacker wants to inject a malicious message that is shown in Figure~\ref{fig:can_malicious_signal}, which contains an identity (ID) field with a value of 0x001, a data length (DLC) field with a value of 0x0, and a cyclical redundancy check (CRC) field with a value of 0x2213.
Note that this malicious message is just an example, and the attacker can craft any valid message as she wishes.
Since the line is always at a recessive state, the attacker only needs to radiate electromagnetic interference when dominant bits need to be injected.
Therefore, the attacker can craft an attacking signal as shown in Figure~\ref{fig:can_malicious_signal}, where the electromagnetic interference corresponds to all 29 dominant bits in this malicious message.
When such an attacking signal is injected into the wires, it first bypasses the transceiver, and then the bypassed injected signal further force the microcontroller to receive dominant bits.
The microcontroller will check the received message, if no error is detected, it will ultimately recognize the malicious message.

We use commercially off-the-shelf electronic devices to build a CAN bus system.
As shown in Figure~\ref{fig:can_attack_real_setup}, a node is connected to one end of two twisted wires.
In the node, a TJA1050 is used as the transceiver, and an ATMEGA328P that is integrated with an MCP2515 CAN controller is used as the microcontroller.
This node is programmed to always listen to the wires.
Moreover, the node is connected to a computer through serial communications so that the received message can be recorded and shown on the computer.
As for the attacking signal, a signal generator is connected to an RF power amplifier, and the amplified signal is radiated by a coil antenna.
In order to inject the attacking signal into the wires effectively and efficiently, the coil antenna is put at around $\SI{5}{\centi\meter}$ above the wires.
Note that this is limited by both local RF equipment regulations and the gain of RF amplifier, but a determined attacker will not be regulated by laws, and she can also increase her attack power by extra cost, thus conducting the attack at a farther distance.
The frequency of the electromagnetic waves is set to be $\SI{22}{\mega\hertz}$ and the amplitude to be $\SI{20}{\volt}$, which has the highest $u$ that is around 0.74 according to preliminary experiments before the attack.
Then, the message injection attack is conducted, and consequently, 3 malicious messages will be successfully recognized every 1000 attacks in 2 seconds, and the success rate is ${0.003}$.

Such a success rate matches our expectations.
Since there are 29 bits to be injected in this message injection, if we regard each bit injection as independent, the expected success rate will be ${0.74^{29} \approx 0.0002}$.
However, as explained in Section~\ref{sec:success_rate_message}, once the first bit injection of several consecutive injections is successful, the success rate for the following injections will be higher, and we approximate the success rate to 1.
As shown in Figure~\ref{fig:can_malicious_signal}, to inject this message, 9 groups of consecutive flips from 1 to 0 are needed, and hence, the expected success rate is around ${0.74^{9} \approx 0.06}$.
A practical result should lie between $0.0002$ and $0.06$, and thus, it is reasonable to obtain a success rate of $0.003$ in practice.
It should be pointed out that in this demonstration, on average there is at least one successful injection per second, and once the malicious message is recognized by the node, it will be executed immediately.

\section{Discussion on Gaining Knowledge}
\label{sec:discussion}

As shown previously, knowing transmitted bits gives an attacker advantage in achieving high success rates of injections.
There are multiple methods of obtaining information about the bits.
For example, a recent work~\cite{rogers2022silently} showed that whatever the state of a CAN bus is, causing two bit errors that are separated by a fixed number of clock cycles can force the bus into an idle state.
Thus, the attacker can use injection attacks to result in such bit errors so as to know the transmitted bits (i.e., 1s).
In addition to actively interfering with the victim system, the attacker can also use existing information about the victim system to figure out what is transmitted.
For example, a preamble of a packet is usually predetermined and published in the protocol, and it is relatively easy to know the transmitted bits in the preamble.
Despite that the payload or checksum may be hard to guess, the attacker can also use a magnetic field probe to listen for electromagnetic leakage from the wires, and the attacker can obtain the bits by analyzing and processing the leakage, which essentially carries information about the bits~\cite{dayanikli2021electromagnetic}.

\section{Related Work}
\label{sec:related_work}

Recent decades have seen an increase in studies about the impacts of electromagnetic interference on electronics: mild interference raises noise level in signals, but powerful interference can destruct the circuits, e.g., flashover, wire melting~\cite{sabath2010classification,giri2020implications,nitsch2004susceptibility}.
In addition to just causing disruption or damage, in the last few years, many studies have started to show that they could use fine-tuned electromagnetic interference to \emph{control} the electronics at a distance.
In this section, we discuss related work about electromagnetic signal injection attacks on digital signals and analog signals, respectively.

\subsection{Electromagnetic Signal Injection Attacks on Digital Signals}

Early work mainly focused on studying how the electromagnetic interference impacts the qualities of digital communications~\cite{mackowski2009influence, li2019statistical, ren2007effects}, but it is more recently that a few studies started to use electromagnetic interference to achieve arbitrary manipulation of digital signals in single-ended lines:
precisely, the attacks can cause a signal receiver to recognize incorrect bits~\cite{selvaraj2018electromagnetic, selvaraj2018intentional}, and with careful synchronization between an attacking signal and a transmitted signal, the success rate of injecting an arbitrary bit injection can reach as high as 100\%~\cite{dayanikli2021electromagnetic}.
In addition, the attacks can be conducted to pulse signals, of which the pulse widths are used to control the actions of actuators.
For example, Selvaraj et al.~\cite{selvaraj2018electromagnetic, selvaraj2018intentional} showed that they can either increase or decrease the pulse widths of the signals so as to manipulate the angles of servos, which are widely used to operate robotic arms or aileron of drones.
Zhang and Rasmussen~\cite{zhang2022detection} showed that similar attacks can also be used to maliciously control the speeds of DC motors, which can be found in insulin pumps and smart locks.
Dayan{\i}kl{\i} et al.~\cite{dayanikli2021electromagnetic, dayanikli2020senact} also demonstrated that they could manipulate the pulse signals that control switches in AC-DC Converters, which play a critical role in the power delivery system of electric vehicles, and the attacks can ultimately lead to short circuit to burn the converter.

Compared with the previous studies, our work pioneers novel attacks on differential signaling, which is supposed to be resistant to external interference.
Moreover, we do not only show a single bit injection but also successfully achieve a message injection, which is not presented in previous work.
It should also be highlighted that by fine-tuning the attack frequency and power, the attacks can still achieve a high success rate of injection, and this helps achieve injections in scenarios where synchronization is difficult.

\subsection{Electromagnetic Signal Injection Attacks on Analog Signals}

While considering the signal injection attacks on analog signals, many studies investigate sensors, whose output signals are essentially analog.
These analog signals are usually at low voltage levels (e.g., several mV), making them more susceptible to electromagnetic interference than digital signals (usually at V levels).
Therefore, it is also relatively easier to attack the analog signals as less attack power is needed.

Rasmussen et al.~\cite{rasmussen2009proximity} identified that electromagnetic signals could induce a current into the audio receiver of the implantable medical devices (IMDs), and later on, Kune et al.~\cite{kune2013ghost} further showed that the electromagnetic signal injection attacks could affect Electrocardiogram (ECG) measurements and make IMDs generate inappropriate defibrillation shocks to heart tissues.
Since the medical devices are safety-critical applications, the attacks can be fatal.
Then, many researchers turn to smartphones~\cite{esteves2018remote, kasmi2015iemi,giechaskiel2019framework, kune2013ghost}. 
In these studies, they demonstrated how to modulate voice commands (e.g., ``OK Google'') onto a high-frequency carrier signal and inject such an attacking signal into smartphones, and the imperfections of electronic components in the smartphones will demodulate the voice commands from the attacking signal, and as such the malicious voice commands will be recognized.
Recently, several studies even showed how to use the electromagnetic waves to inject fake touches onto the touchscreens of the smartphones so as to realize manipulation such as installing malicious applications~\cite{wang2022ghosttouch, shan2022ghosttouch,jiang2022wight}.
Since the electromagnetic waves are neither visible nor audible, the attacks can maliciously control the smartphone quietly, without being noticed by the users.
In addition, researchers also showed how to maliciously control temperature measurements of thermometers, which are widely used for temperature monitoring applications such as baby incubators and nuclear reactors~\cite{tu2019trick, sp20Zhang}.
K{\"o}hler et al.~\cite{kohler2021signal} demonstrated that they could use electromagnetic signals to manipulate frames captured by image sensors, which are widely used for logistics applications to military systems.

\section{Conclusion}
\label{sec:conclusion}

Despite the fact that differential signaling was proposed for making communication cables more immune to external interference, in this work, we show that electromagnetic signal injection attacks can inject arbitrary information into a differential signaling system.
Because of the input asymmetry and nonlinearities of the subtractor, the rejection ability of the differential signaling technique is not sufficiently good for high-frequency signals to prevent attackers from successfully injecting adversarial signals.
Moreover, in the receiver, the ESD circuit's rectification plus the buffer circuit's net charge accumulation results in high-frequency signals ultimately being incorrectly detected as either 0's or 1's depending on the frequency.
Our experiments have demonstrated the attack principles, and how to properly choose the frequency and the power of the attacking signal, in order to achieve successful injection.
We analyze the success rate of injection of more complicated bitstrings, taking into account any knowledge that the attacker might have about the existing data transmissions in the cable. We show how this knowledge and the choice of attacking signals will affect the success rate.
This analysis can also be used defensively by system designers who want to evaluate the security of their own systems, and are able to change the components and data modulation scheme to minimize adversarial success. Finally, we demonstrate arbitrary message injection into a CAN bus, allowing an attacker to dictate the actions of the victim system.

\balance
\bibliographystyle{IEEEtranS}
\bibliography{mybib}
\end{document}